\documentclass[twocolumn]{aastex701}

\usepackage{amsmath}
\usepackage{mhchem}

\newcommand{\spectuner}{\texttt{spectuner}}
\newcommand{\astrodendro}{\texttt{astrodendro}}
\newcommand{\radmc}{\texttt{RADMC-3D}}

\newcommand{\kms}{km\,s$\rm ^{-1}$}

\def \ch3cn{CH$_3$CN}

\begin{document}

\title{The ALMA-QUARKS survey: Investigating Thermal Feedback of Massive Protostars in Hot Molecular Cores}

\correspondingauthor{Dezhao Meng, Tie Liu, Jarken Esimbek}

\author[orcid=0009-0000-5764-8527]{Dezhao Meng}
\affiliation{State Key Laboratory of Radio Astronomy and Technology, Xinjiang Astronomical Observatory, Chinese Academy of Sciences, \\150 Science 1-Street, Urumqi, Xinjiang, 830011, People’s Republic of China}
\affiliation{University of Chinese Academy of Sciences, Beijing 100080,
People’s Republic of China}
\affiliation{State Key Laboratory of Radio Astronomy and Technology, Shanghai Astronomical Observatory, Chinese Academy of Sciences, \\
80 Nandan Road, Shanghai 200030, People's Republic of China}
\email[show]{mengdezhao@xao.ac.cn}  

\author[orcid=0000-0002-5286-2564]{Tie Liu}
\affiliation{State Key Laboratory of Radio Astronomy and Technology, Shanghai Astronomical Observatory, Chinese Academy of Sciences, \\
80 Nandan Road, Shanghai 200030, People's Republic of China}
\affiliation{University of Chinese Academy of Sciences, Beijing 100080,
People’s Republic of China}
\email[show]{liutie@shao.ac.cn} 

\author[orcid=0000-0003-4910-1390]{Jarken Esimbek}
\affiliation{State Key Laboratory of Radio Astronomy and Technology, Xinjiang Astronomical Observatory, Chinese Academy of Sciences, \\150 Science 1-Street, Urumqi, Xinjiang, 830011, People’s Republic of China}
\affiliation{University of Chinese Academy of Sciences, Beijing 100080,
People’s Republic of China}
\affiliation{Xinjiang Key Laboratory of Radio Astrophysics, Urumqi 830011, People’s Republic of China}
\email[show]{jarken@xao.ac.cn}

\author[orcid=0000-0002-7716-1094]{Yisheng Qiu}
\affiliation{Research Center for Computational Earth and Space Science, Zhejiang Laboratory, Hangzhou 311121, People’s Republic of China}
\email{yishengq@zhejianglab.org}

\author[orcid=0000-0002-9600-1846]{Jixing Ge}
\affiliation{State Key Laboratory of Radio Astronomy and Technology, Xinjiang Astronomical Observatory, Chinese Academy of Sciences, \\150 Science 1-Street, Urumqi, Xinjiang, 830011, People’s Republic of China}
\affiliation{University of Chinese Academy of Sciences, Beijing 100080,
People’s Republic of China}
\email{gejixing666@163.com}

\author[0000-0001-5175-1777]{Neal J. Evans II}
\affiliation{Department of Astronomy, The University of Texas at Austin, 2515 Speedway, Stop C1400, Austin, TX 78712-1205, USA}
\email{nje@astro.as.utexas.edu}

\author[0000-0002-9569-9234]{Aina Palau}
\affiliation{Instituto de Radioastronom\'ia y Astrof\'isica, Universidad Nacional Aut\'onoma de M\'exico, Antigua Carretera a P\'atzcuaro 8701, Ex-Hda. San Jos\'e de la Huerta, Morelia, 58089, Michoac\'an, M\'exico}
\email{a.palau@irya.unam.mx}

\author[0000-0003-1649-7958]{Guido Garay}
\affiliation{Chinese Academy of Sciences South America Center for Astronomy, National Astronomical Observatories, CAS, Beijing 100101, People’s Republic of China}
\affiliation{Departamento de Astronom\'ia, Universidad de Chile, Camino el Observatorio 1515, Las Condes, Santiago, Chile}
\email{guido@das.uchile.cl}

\author[0000-0002-6622-8396]{Paul F. Goldsmith}
\affiliation{Jet Propulsion Laboratory, California Institute of Technology, 4800 Oak Grove Drive, Pasadena, CA 91109, USA}
\email{paul.f.goldsmith@jpl.nasa.gov}

\author[0000-0001-5950-1932]{Fengwei Xu}
\affiliation{Max Planck Institute for Astronomy, K\"{o}nigstuhl 17, 69117 Heidelberg, Germany}
\email{fengweilookuper@gmail.com}

\author[0000-0002-8697-9808]{Sami Dib}
\affiliation{Max Planck Institute for Astronomy, K\"{o}nigstuhl 17, 69117 Heidelberg, Germany}
\email{sami.dib@gmail.com}

\author[0000-0003-3119-2087]{Jeong-Eun Lee}
\affiliation{Department of Physics and Astronomy, SNU Astronomy Research Center, Seoul National University, 1 Gwanak-ro, Gwanak-gu, Seoul 08826, Republic of Korea}
\email{lee.jeongeun@snu.ac.kr}

\author[0000-0003-2300-8200]{Amelia M.\ Stutz}
\affiliation{Departamento de Astronom\'{i}a, Universidad de Concepci\'{o}n,Casilla 160-C, Concepci\'{o}n, Chile}
\email{amelia.stutz@gmail.com}

\author[0000-0002-4154-4309]{Xindi Tang}
\affiliation{State Key Laboratory of Radio Astronomy and Technology, Xinjiang Astronomical Observatory, Chinese Academy of Sciences, \\150 Science 1-Street, Urumqi, Xinjiang, 830011, People’s Republic of China}
\affiliation{University of Chinese Academy of Sciences, Beijing 100080,
People’s Republic of China}
\affiliation{Xinjiang Key Laboratory of Radio Astrophysics, Urumqi 830011, People’s Republic of China}
\email{tangxindi@xao.ac.cn}

\author[0000-0001-7573-0145]{Xiaofeng Mai}
\affiliation{State Key Laboratory of Radio Astronomy and Technology, Shanghai Astronomical Observatory, Chinese Academy of Sciences, \\
80 Nandan Road, Shanghai 200030, People's Republic of China}
\email{maixf@shao.ac.cn}

\author[orcid=0000-0001-7817-1975,sname='Y.-K. Zhang']{Yan-Kun Zhang}
\affiliation{State Key Laboratory of Radio Astronomy and Technology, Shanghai Astronomical Observatory, Chinese Academy of Sciences, \\
80 Nandan Road, Shanghai 200030, People's Republic of China}
\email{zhangyankun@shao.ac.cn}

\author[0000-0001-9822-7817]{Wenyu Jiao}
\affiliation{State Key Laboratory of Radio Astronomy and Technology, Shanghai Astronomical Observatory, Chinese Academy of Sciences, \\
80 Nandan Road, Shanghai 200030, People's Republic of China}
\email[hide]{wenyujiao@shao.ac.cn}

\author[0009-0000-9090-9960]{Jia-Hang Zou}
\affiliation{State Key Laboratory of Radio Astronomy and Technology, Shanghai Astronomical Observatory, Chinese Academy of Sciences, \\
80 Nandan Road, Shanghai 200030, People's Republic of China}
\affiliation{School of Physics and Astronomy, Yunnan University, Kunming 650091, People’s Republic of China}
\email{zojh1000@163.com}

\author[0000-0002-9574-8454]{Leonardo Bronfman}
\affiliation{Departamento de Astronom\'ia, Universidad de Chile, Las Condes, 7591245 Santiago, Chile}
\email{leo@das.uchile.cl}

\author[0000-0001-7151-0882]{Swagat R. Das} 
\affiliation{Departamento de Astronom\'ia, Universidad de Chile, Las Condes, 7591245 Santiago, Chile}
\email{swagat@das.uchile.cl}

\author[0000-0003-1602-6849]{Prasanta Gorai}
\affiliation{Rosseland Centre for Solar Physics, University of Oslo, PO Box 1029 Blindern, 0315 Oslo, Norway}
\affiliation{Institute of Theoretical Astrophysics, University of Oslo, PO Box 1029 Blindern, 0315 Oslo, Norway}
\email{prasanta.astro@gmail.com}

\author[0009-0004-2927-239X]{Jian-Wen Zhou}
\affiliation{Max-Planck-Institut für Radioastronomie, Auf dem H\"{u}gel 69, D-53121 Bonn, Germany}
\email{jwzhou@mpifr-bonn.mpg.de}

\author[orcid=0000-0002-8586-6721]{Pablo Garc\'ia}
\affiliation{Chinese Academy of Sciences South America Center for Astronomy, National Astronomical Observatories, CAS, Beijing 100101, People’s Republic of China}
\affiliation{Instituto de Astronom\'ia, Universidad Cat\'olica del Norte, Av. Angamos 0610, Antofagasta, Chile}
\email{astro.pablo.garcia@gmail.com}

\author[0000-0002-5310-4212]{L. Viktor T\'oth}
\affiliation{Institute of Physics and Astronomy, E\"otv\"os Lor\'and University, P\'azm\'any P\'eter s\'et\'any 1/A, H-1117 Budapest, Hungary}
\affiliation{Faculty of Science and Technology, University of Debrecen, H-4032 Debrecen, Hungary}
\email{toth.laszlo.viktor@ttk.elte.hu}

\author[0000-0003-4603-7119]{Sheng-Yuan Liu}
\affiliation{Institute of Astronomy and Astrophysics, Academia Sinica, 11F of ASMAB, AS/NTU No.1, Sec. 4, Roosevelt Road, Taipei 10617,
Taiwan}
\email{syliu@asiaa.sinica.edu.tw}

\author[0000-0002-7125-7685]{Patricio Sanhueza}
\affiliation{Department of Astronomy, School of Science, The University of Tokyo, 7-3-1 Hongo, Bunkyo-ku, Tokyo 113-0033, Japan}
\email{patosanhueza@gmail.com}

\author[0000-0002-3179-6334]{Chang Won Lee}
\affiliation{Korea Astronomy and Space Science Institute, 776 Daedeokdaero, Yuseong-gu, Daejeon 34055, Republic of Korea}
\affiliation{University of Science and Technology, Korea (UST), 217 Gajeong-ro, Yuseong-gu, Daejeon 34113, Republic of Korea}
\email{cwl@kasi.re.kr}

\author[0000-0003-2412-7092]{Kee-Tae Kim}
\affiliation{Korea Astronomy and Space Science Institute, 776 Daedeokdaero, Yuseong-gu, Daejeon 34055, Republic of Korea}
\affiliation{University of Science and Technology, Korea (UST), 217 Gajeong-ro, Yuseong-gu, Daejeon 34113, Republic of Korea}
\email{ktkim@kasi.re.kr}

\author[0000-0003-0356-818X]{Jianjun Zhou}
\affiliation{State Key Laboratory of Radio Astronomy and Technology, Xinjiang Astronomical Observatory, Chinese Academy of Sciences, \\150 Science 1-Street, Urumqi, Xinjiang, 830011, People’s Republic of China}
\affiliation{University of Chinese Academy of Sciences, Beijing 100080,
People’s Republic of China}
\affiliation{Xinjiang Key Laboratory of Radio Astrophysics, Urumqi 830011, People’s Republic of China}
\email{zhoujj@xao.ac.cn}

\author[0000-0003-0933-7112]{Gang Wu}
\affiliation{State Key Laboratory of Radio Astronomy and Technology, Xinjiang Astronomical Observatory, Chinese Academy of Sciences, \\150 Science 1-Street, Urumqi, Xinjiang, 830011, People’s Republic of China}
\affiliation{University of Chinese Academy of Sciences, Beijing 100080,
People’s Republic of China}
\affiliation{Xinjiang Key Laboratory of Radio Astrophysics, Urumqi 830011, People’s Republic of China}
\email{wug@xao.ac.cn}

\author[0000-0001-5494-6238]{Dalei Li}
\affiliation{State Key Laboratory of Radio Astronomy and Technology, Xinjiang Astronomical Observatory, Chinese Academy of Sciences, \\150 Science 1-Street, Urumqi, Xinjiang, 830011, People’s Republic of China}
\affiliation{University of Chinese Academy of Sciences, Beijing 100080,
People’s Republic of China}
\affiliation{Xinjiang Key Laboratory of Radio Astrophysics, Urumqi 830011, People’s Republic of China}
\email{lidalei@xao.ac.cn}

\author[0000-0002-8760-8988]{Yuxin He}
\affiliation{State Key Laboratory of Radio Astronomy and Technology, Xinjiang Astronomical Observatory, Chinese Academy of Sciences, \\150 Science 1-Street, Urumqi, Xinjiang, 830011, People’s Republic of China}
\affiliation{University of Chinese Academy of Sciences, Beijing 100080,
People’s Republic of China}
\affiliation{Xinjiang Key Laboratory of Radio Astrophysics, Urumqi 830011, People’s Republic of China}
\email{heyuxin@xao.ac.cn}

\author[0009-0004-6159-5375]{Dongting Yang}
\affiliation{School of Physics and Astronomy, Yunnan University, Kunming 650091, People’s Republic of China}
\email{dongting@mail.ynu.edu.cn}

\author[0000-0002-9875-7436]{James O. Chibueze}
\affiliation{UNISA Centre for Astrophysics and Space Sciences, College of Science, Engineering and Technology, University of South Africa, Cnr Christian de Wet Rd and Pioneer Avenue, Florida Park, 1709, Roodepoort, South Africa}
\affiliation{Department of Physics and Astronomy, Faculty of Physical Sciences, University of Nigeria, Carver Building, 1 University Road, Nsukka 410001, Nigeria}
\email{james.chibueze@gmail.com}

\author[0000-0001-8315-4248]{Xunchuan Liu}
\affiliation{State Key Laboratory of Radio Astronomy and Technology, Shanghai Astronomical Observatory, Chinese Academy of Sciences, \\
80 Nandan Road, Shanghai 200030, People's Republic of China}
\email{liuxunchuan001@gmail.com}

\author[]{Lei Zhu}
\affiliation{Chinese Academy of Sciences South America Center for Astronomy, National Astronomical Observatories, CAS, Beijing 100101, People’s Republic of China}
\email{lzhupku@gmail.com}


\begin{abstract}
We identify a sample of 83 spatially resolved hot molecular cores (HMCs) in the QUARKS survey, aiming at investigating thermal feedback from massive stars.
Using CH$_3$CN\,(12--11) line emission together with 1.3\,mm continuum data we derive the radial temperature, volume density and \ch3cn{} abundance profiles for the 83 HMCs.
Based on the envelope temperature and density profiles, we compute the luminosities of the embedded massive protostars with \radmc{} radiation transfer model.
The derived luminosities are comparable (within $\sim1$ dex) to the bolometric luminosities of their natal clumps and show strong correlations with several core-scale properties, including the HMC mass
($Log[ M_\mathrm{env}] = 1.01\,Log [L_\star] - 4.80$), the inner core radius (the flat radius of Plummer-like volume density profile)
($Log[a] = 0.46\,Log[L_\star] + 0.52$) and the central density $ (Log[n_c] = -0.55 Log[L_\star] +10.47) $.
These empirical relations provide useful observational constraints for physical models of protostellar objects.
Importantly, we find a strong positive correlation between the massive protostellar luminosity and the local thermal Jeans mass.
The derived Jeans masses, $M_\mathrm{Jeans}$, exceed the HMC masses $M_\mathrm{env}$, with the average $M_\mathrm{Jeans}$ being two times larger than the average $M_\mathrm{env}$. 
This provides observational evidence that thermal feedback from massive protostars can effectively suppress further fragmentation of HMCs, thereby promoting massive star formation.
In addition, the positive correlation between massive protostellar luminosity and natal clump mass suggests that more massive clumps preferentially host more luminous protostars, leading to stronger thermal feedback.
\end{abstract}

\section{INTRODUCTION}

\setcounter{footnote}{0}

Newly formed massive protostars are capable of efficiently heating their surrounding gas through intense radiative output, 
establishing temperature structures that extend over their natal envelopes.
This thermal feedback is expected to play a crucial role in the formation of high-mass stars by regulating fragmentation, accretion, and the physical states of their natal cores \citep{2014prpl.conf..243K}, thus potentially affecting the shape of the Initial Mass Function \citep{Palau2024}.

Over the past decade, the importance of thermal feedback has been extensively demonstrated in numerical simulations of massive star formation \citep{2014prpl.conf..243K}. 
Radiation-hydrodynamic models suggest that protostellar heating significantly suppresses fragmentation compared to isothermal or barotropic cases, resulting in fewer but more massive cores and further modifying stellar mass spectra
\citep[e.g.,][]{Krumholz2006,Krumholz2007,Krumholz2008,Bate2009,Bate2012,Krumholz2010,Krumholz2011,Krumholz2012,Krumholz2016,Offner2009,Urban2010,Myers2013,Li2018,Guszejnov2016,Guszejnov2022}.
This has been tentatively observed in individual massive star forming regions such as W51 (e.g., \citealt{Tang2022}), and
is a natural explanation for the different properties of minimum masses and binary separations in the substellar regime in nearby
clouds (e.g., \citealt{Palau2024,Bouy2026}).
In addition, numerous studies using different temperature tracers have identified radially decreasing, power-law temperature structures attributed to stellar thermal feedback across a wide range of spatial scales \citep[e.g.,][]{Longmore2011,Hatchell2013,Sicilia2013,Lu2014,Giannetti2017,Tang2018a,Tang2018b,Tang2021,Gieser2019,Gieser2021,Gieser2022,Gieser2023,Li2020,Lin2022,Wang2023,Estalella2024,Jeff2024,Zhao2024,Motte2025}. However, most of these studies focus on individual sources or small samples. Because they employ inconsistent observational setups—including different angular resolutions and temperature tracers—it remains challenging to combine existing measurements into a coherent framework and to draw a robust conclusion on the role of thermal feedback in massive star formation.

With the advent of ALMA, the greatly improved angular resolution and sensitivity now enable detailed measurements of the density and temperature structures of protostellar envelopes. These capabilities provide powerful means to directly probe the physical conditions regulated by protostellar radiative heating.
In this paper, leveraging the Querying Underlying mechanisms of massive star formation with ALMA-Resolved gas Kinematics and Structures (QUARKS, \citealt{quarks1,Mai2024,Yang2024,quarks2,quarks3,2025ApJ...995..193Y,Zou2025,Zhang2025,Meng2025,2025arXiv251210298Y}) survey, we investigate the temperature and density structures of a large sample of hot molecular cores (HMCs or hot cores) to evaluate the thermal feedback from the embedded massive protostars. 

\section{THE SAMPLE and ALMA OBSERVATIONS}

Our sample consists of 58 massive star-forming clumps selected from the QUARKS survey, whose basic parameters are presented in \citet{Liu2020} and \citet{quarks1}. The source selection criterion is that the spatial extent of the CH$_3$CN\,(12–-11, K=0–4) moment\,0 emission exceeds four times the synthesized beam, ensuring that the core structure well resolved.
These spatially extended CH$_3$CN-emitting structures trace HMCs, which are recognized as a long-standing evolutionary phase in the formation of massive protostars \citep{Meng2025}.
The distances to these sources range from 1.3 to 12.9 kpc, with a median value of 4.3 kpc.
The masses of the natal clumps associated with these sources span from 250 to $5.0\times10^{4}\,M_\odot$, with a median mass of $3.5\times10^{3}\,M_\odot$.
The bolometric luminosities range from $1\times10^{3}$ to $5\times10^{6}\,L_\odot$, with a median value of $1\times10^{5}\,L_\odot$.
The entire set of values can be found in \citet{Liu2020}.
The methods used to derive the clump masses and bolometric luminosities are described in \citet{Urquhart2018}.

These sources were obsereved in the QUARKS survey (Project ID: 2021.1.00095.S; PIs: Lei Zhu,
Guido Garay and Tie Liu) with ALMA at Band 6 (1.3 mm). For each source, observations were conducted using the Atacama Compact 7-m Array, ALMA 12-m array in C-2 and C-5 configurations. The combined data from these three configurations provide a angular resolution of $\sim$0.3\arcsec{}, a continuum rms noise $\sigma_{\mathrm{cont}}$ in a wide range from 0.1 to several mJy\,beam$^{-1}$ for different sources, and a Maximum Recoverable Scale (MRS) of $\sim$27\arcsec{}. The typical rms noise $\sigma_{\mathrm{line}}$ for lines is $\sim$3 mJy\,beam$^{-1}$ per 0.976 MHz channel ($\sim$1.3 \kms). More detailed information on observation and data reduction can be found in \citet{quarks1}, \citet{quarks2}, and \citet{quarks3}. In this work, we use the 1.3~mm continuum data, as well as line data of \ch3cn\,(12--11) and H30$\alpha$.

\section{RESULTS}
\begin{figure*}[ht!]
    \centering
    \includegraphics[width=18cm]{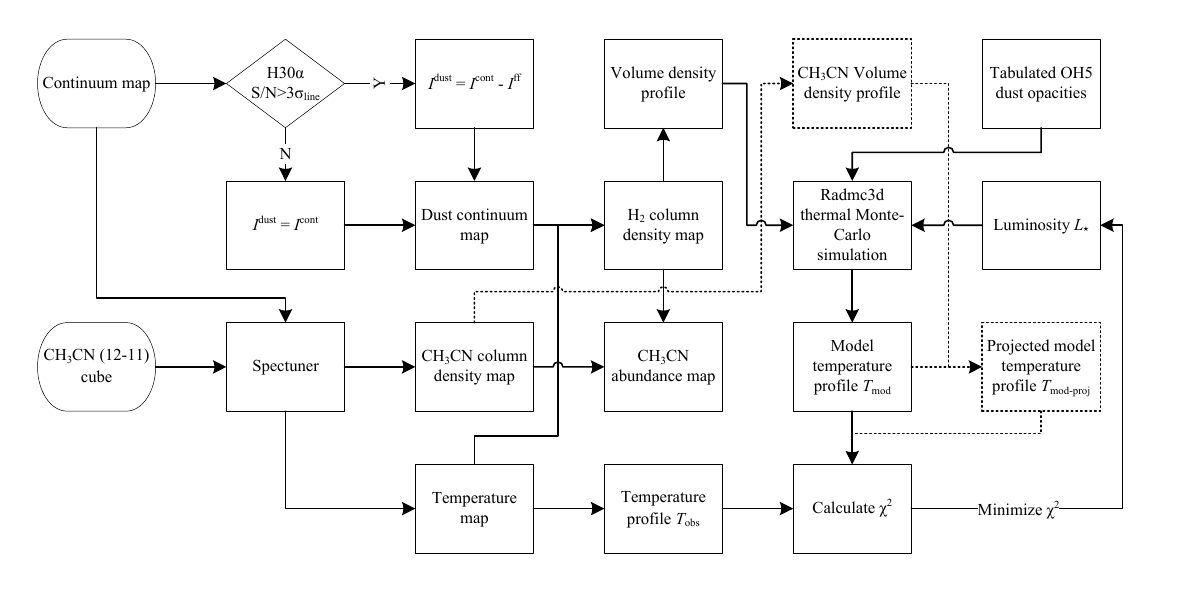}
    \caption{Overall workflow illustrating the calculation of most physical quantities in this paper; detailed procedures are provided in the main text.  The dashed lines indicate projection-corrected \radmc{} model, which is realized in Appendix \ref{Projection-corrected}.
    }
    \label{fig:workflow}
\end{figure*}

In this section, we derive temperature and column density maps for all sources in our sample. Based on these maps, we further calculate a set of physical properties for the HMCs, which are used in the subsequent analysis and discussion. The overall procedure follows the workflow illustrated in Figure~\ref{fig:workflow}.

\begin{figure*}[ht!]
    \centering
    \includegraphics[width=18cm]{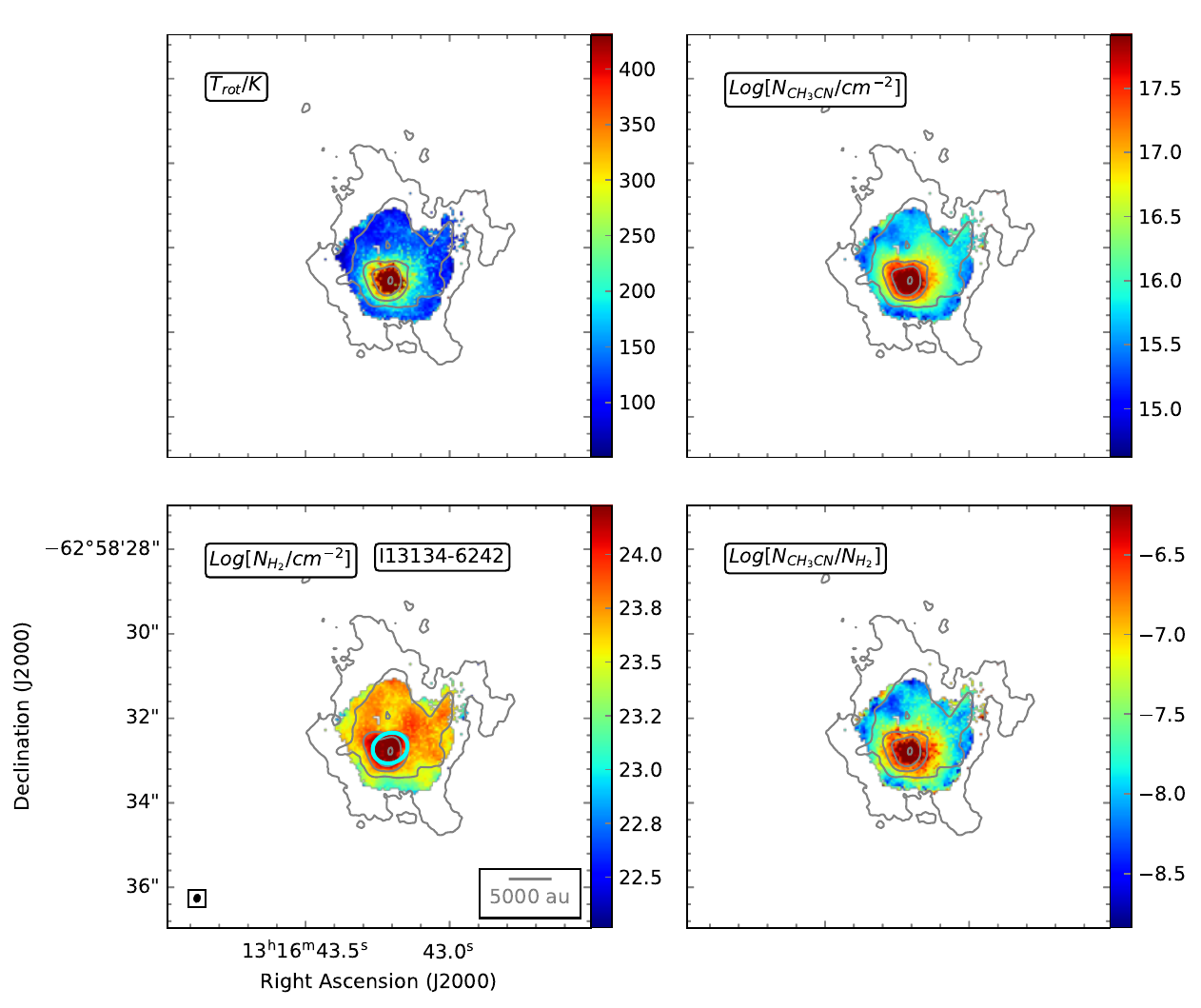}
    \caption{Example images for field I13134-6242 showing the spatial distribution of different parameters. The corresponding physical quantities are labeled in the upper-left corner of each panel. Upper panels: rotational temperature map (left panel) and \ch3cn{} column density map (right panel) produced by simultaneously fitting the multi-transitions of \ch3cn\,(12--11) pixel-by-pixel using \spectuner{}. Lower-left panel: H$_2$ column density map derived from \ch3cn{} rotational temperature map and 1.3~mm continuum map. The cyan ellipse outlines the I13134-6242-HC1 structure identified by \astrodendro{}. The synthesized beam is shown in the lower-left corner, and the scale bar is indicated in the lower-right corner. Lower-right panel: \ch3cn{} abundance map. The gray contours for the four panels represent the 1.3~mm continuum. The contour levels were plotted from 3$\sigma_{\mathrm{cont}}$ (1.5 mJy\,beam$^{-1}$) to 0.95 times the peak value, with 5 logarithmically spaced contours between these values.
    The similar images for the remaining sources are displayed in the complete figure set, which is available in the online journal.  
    }
    \label{fig:image_example}
\end{figure*}

\begin{figure*}[ht!]
    \centering
    \includegraphics[scale=0.8]{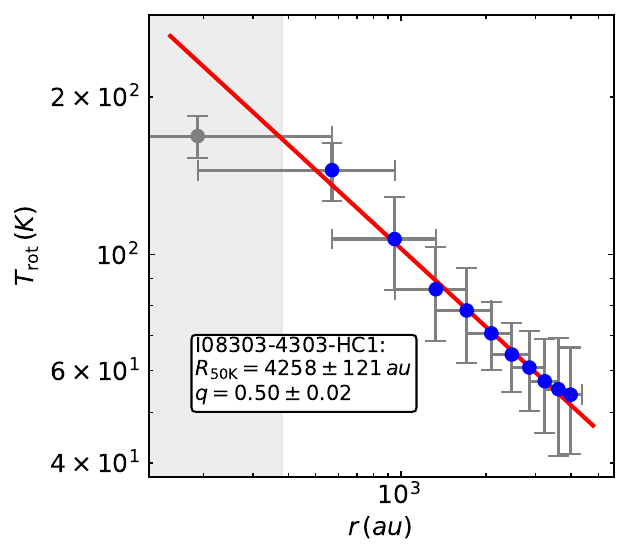} \hspace{0cm}
    \includegraphics[scale=0.8]{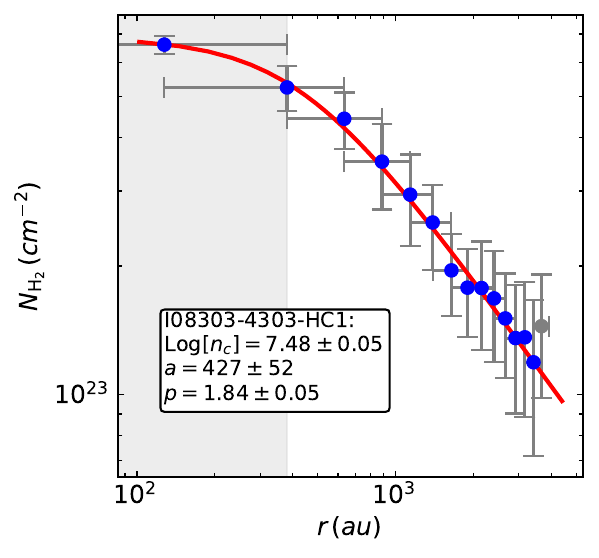} \hfill
    \includegraphics[scale=0.8]{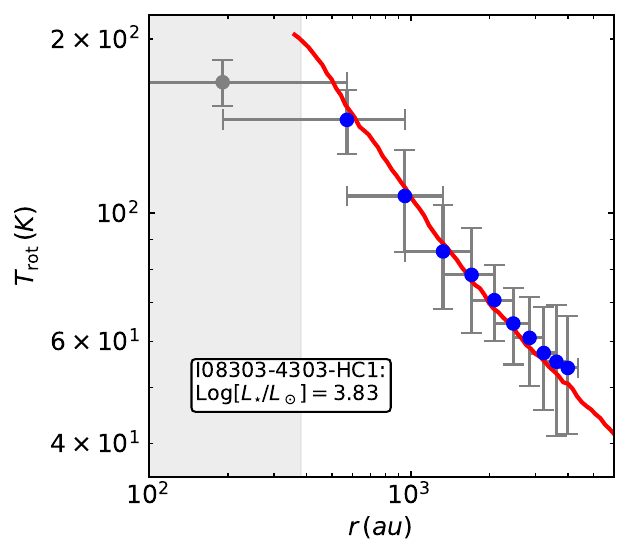} \hspace{0cm}
    \includegraphics[scale=0.8]{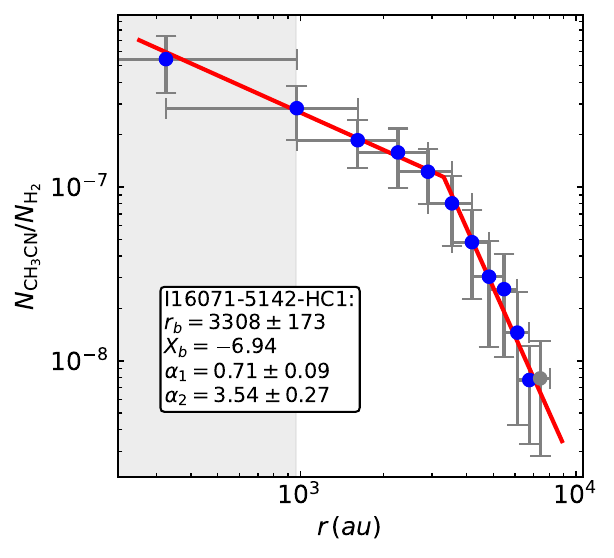}
    \caption{Representative example for the fitting results for the radial profile of different parameters. Upper-left and lower-left panels display the binned radial temperature profile derived from the \ch3cn{} rotational temperature map, but the different models (upper: Equation~\ref{eq: T_pow-law}; lower: \radmc{} model) are used to fit both in order to get different physical quantities. The upper-right panel and lower-right panel show the binned radial H$_2$ column density profile and \ch3cn{} abundance profile, which were derived from the H$_2$ column density map and \ch3cn{} abundance map shown in the lower-left and lower-right panels of Figure~\ref{fig:image_example}, respectively. The data points used for the radial profile fit are shown in blue, while gray points are excluded from the model fits. The red solid lines show the best-fitted models. All physical parameters derived from the fits are shown in the legend of each panel. The inner unresolved region is shown as a grey-shaded area. 
    The similar images for the remaining sources are displayed in the complete figure set, which is available in the online journal.
    }
    \label{fig: profile}
\end{figure*}

\begin{figure*}[ht!]
    \centering
    \includegraphics[width=18cm]{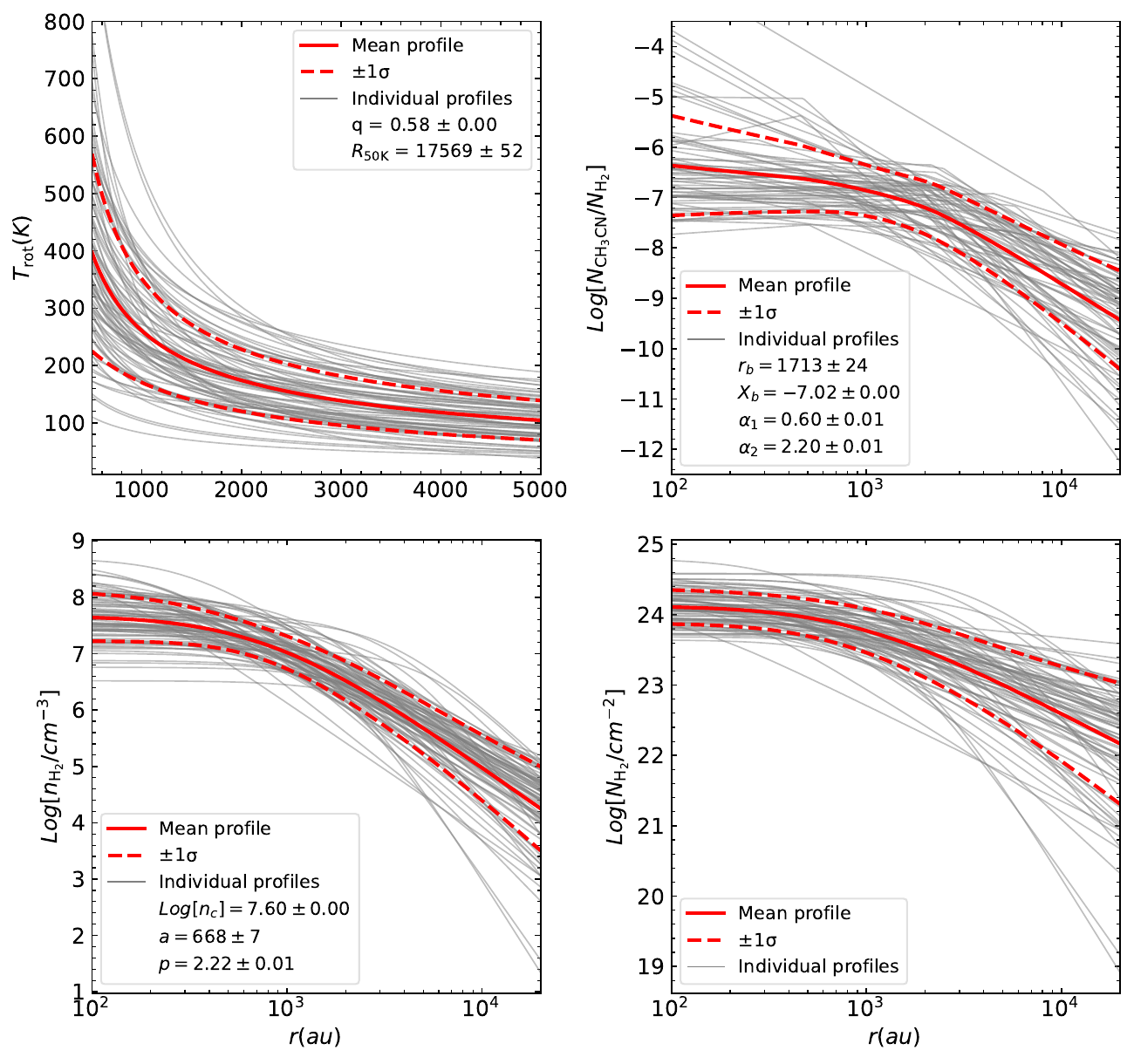}
    \caption{The radial profiles of temperature (upper-left), \ch3cn{} abundance (upper-right), H$_2$ volume density (lower-left) and column density (lower-right)  for all HMCs.  The thin gray lines are the profiles for individual HMCs. The solid red lines present the averaged profile, and the dashed red line indicate the 1\,$\sigma$ uncertainty ranges of the mean profiles. Note that we exclude the temperature profiles of I16348-4654-HC1 and I18056-1952-HC1, as their unusually high temperatures significantly deviate from those of the other HMCs and are likely unreliable.
    The average temperature, \ch3cn{} abundance and volume density profile can be fitted with Equation~\ref{eq: T_pow-law}, \ref{eq:abundance} and \ref{eq:volume density}, respectively. The best-fit parameters are labeled in the corresponding panels.
    }
    \label{fig:mean_profile}
\end{figure*}

\subsection{The temperature and column density maps}

\ch3cn\,(12--11) is capable of tracing dense gas in the immediate vicinity of protostars and serves as an excellent thermometer in high-density regions \citep{Loren1984}. 
We therefore use the multiple K-ladder transitions of \ch3cn\,(12--11) to derive the temperature distribution of protostellar envelopes. 
\spectuner{}\footnote{\url{https://github.com/yqiuu/spectuner}} \citep{Qiu2025,Qiu2026} is a Python package for automated spectral line fitting, which employs the spectral line model from \citet{2017A&A...598A...7M} and utilizes deep reinforcement learning. The spectra are fitted using the SLSQP optimization algorithm, with initial guesses provided by a deep neural network described in \citet{Qiu2026}. The neural network was trained on real ALMA observations, including data from the ALMA Three-millimeter Observations of Massive Star-forming regions (ATOMS) survey \citep{Liu2020}. 
In this work, \spectuner{} is used to derive the rotational temperature ($T_\mathrm{rot}$) maps by simultaneously fitting the spectra of the multi-transition \ch3cn\,(12--11, K=0--8, $E_u/k_B = 69-526\,K$, where $E_u$ is the upper-level energy and $k_B$ is the Boltzmann constant) emission on a pixel-by-pixel basis.
The 1.3~mm continuum maps are incorporated into \spectuner{} to estimate the background temperature for the spectral-line models.
In the \spectuner{} configuration, the beam-filling factor, column density, temperature, velocity, and line width are treated as free parameters during the fitting process. 
We adopt the peak matching loss function for the optimization, which was developed by \citet{Qiu2025} and is particularly suitable when line blending is significant. 
Finally, we apply a mask to the column density, temperature, velocity, and line width maps produced by \spectuner{}, selecting only pixels with $S_3^i>0.6$ and $S_{\text {tot }}^i>2.4$ (the definition of $S_3^i$ and $S_{\text {tot }}^i$ can be found in \citealt{Qiu2025}). These criteria indicate a good quality of the spectral fits, while all remaining pixels are set to NaN. 
Because this work only focuses on temperature and density maps, an example for the rotational temperature and \ch3cn\ column density maps is shown in the upper panels of Figure \ref{fig:image_example}.
Under the assumption of local thermal equilibrium (LTE), the gas kinetic temperature can be approximated by the rotational temperature, i.e., $T_{\mathrm{kin}} \approx T_{\mathrm{rot}}$. This approximation is generally valid because population exchange among K-ladders is primarily driven by collisions.

To compute H$_2$ column density maps for these fields, we first assume dust temperature $T_{\mathrm{dust}} \approx T_{\mathrm{kin}}$.
This is a reasonable approximation in the high-density ($n_\mathrm{H_2} \ge 10^5\, \mathrm{cm}^{-3}$, \citealt{Goldsmith2001,Young2004,Ivlev2019}) environments traced by \ch3cn\,(12--11), where the dust and gas are expected to be thermally coupled. 
Secondly, we calculate the peak continuum optical depth for each hot core (identified in Section \ref{sec: The identification of internally heated hot core}) using the following equation \citep{Frau2010}:
\begin{equation}
\tau_\nu^{\text {cont}}=-\ln \left(1-\frac{I_\nu^{\text {cont }}}{\Omega B_\nu(T_\mathrm{dust})}\right) ,
\end{equation}
where $I_\nu^{\text {cont }}$ is flux density in Jy\,beam$^{-1}$, $\Omega$ is the beam solid angle in sr (i.e. $\Omega=\frac{\pi}{4 \ln (2)} \theta_{\mathrm{bmaj}}\theta_{\mathrm{bmin}}$, $\theta_{\mathrm{bmaj}}$ and $\theta_{\mathrm{bmin}}$ are the FWHM of the major and minor axis of the beam), and $B_\nu(T_\mathrm{dust})$ is the Planck function. The calculated continuum optical depth, $\tau_\nu^{\mathrm{cont}}$, is $\ll 1$ at most peak continuum positions, indicating that the optically thin assumption for 1.3 mm continuum emission is valid in our sample.
Thirdly, for hot cores with detected H30$\alpha$ emission (called ``dust-ff" hot cores; details can be found in Section \ref{sec: The identification of internally heated hot core}), we subtract the free--free contribution estimated through the following equation \citep{Wilson2009}:
\begin{equation}
I_\nu^{\text {ff}}= 5.241 \times 10^{-5}\nu^{-1} T_e\left(1+N_{\mathrm{He}} / N_{\mathrm{H}}\right) \left\langle g_{\mathrm{ff}}\right\rangle \sum I_{\mathrm{H} 30\alpha} \Delta \mathrm{v},
\end{equation}
where the Gaunt factor for free--free transitions is given by $\left\langle g_{\mathrm{ff}}\right\rangle = \ln \left[4.955 \times 10^{-2} \nu^{-1}\right]+1.5 \ln \left(T_{\mathrm{e}}\right)$
, the electron temperature $T_{\mathrm e}$ in K is adopted from \citet{Zhang2023}, helium-to-hydrogen number ratio $N_{\mathrm{He}} / N_{\mathrm{H}} =0.08$ \citep{Churchwell1974,Lichten1979}, $\nu(\sim\,232\,\mathrm{GHz})$ denotes the H30$\alpha$ central frequency in GHz, and the velocity-integrated intensity $\sum I_{\mathrm{H}30\alpha}\,\Delta \mathrm{v}$ is measured from the moment~0 map in units of Jy beam$^{-1}$ \kms. Then, the dust continuum maps are estimated through the $I_\nu^{\text {dust}} = I_\nu^{\text {cont}} - I_\nu^{\text {ff}}$. Finally, combined with $T_{\mathrm{dust}}$ ($=T_{\mathrm{rot}}$) and $I_\nu^{\text {dust}}$ maps, H$_2$ column density maps can be calculated pixel-by-pixel using the following equation: 
\begin{equation}
N_{\mathrm{H}_2} =
\mathcal{R}\,
\frac{I_\nu^{\text {dust}}}
{\Omega\,\mu\,m_{\mathrm H}\,\kappa_\nu\,B_\nu(T_{\mathrm{dust}})} ,
\end{equation}
where $\mu = 2.81$ is the molecular weight per hydrogen molecule, $m_{\mathrm H}$ is the mass of a hydrogen atom, gas-to-dust mass ratio $\mathcal{R} = 100$ and dust opacity $\kappa_\nu = 1\,\mathrm{cm}^2\,\mathrm{g}^{-1}$ at $\nu \sim 230\, \mathrm{GHz}$, which is interpolated from the tabulated values in \citet{Ossenkopf&Henning1994}, assuming grains with thin ice mantles, a size distribution following \citet{Mathis1977}, and a typical gas density of $10^{6}\,\mathrm{cm}^{-3}$. An example for H$_2$ column density maps is displayed in the lower-left panel of Figure \ref{fig:image_example}. Furthermore, the \ch3cn\ abundance maps are derived from the ratio of the \ch3cn\ column density maps to the H$_2$ column density maps and the lower-right panel in Figure~\ref{fig:image_example} displays the corresponding \ch3cn\ abundance map.

\subsection{The identification of internally heated hot cores}
\label{sec: The identification of internally heated hot core}

To extract the core structures where temperature structures are resolved very well, we first use \astrodendro{}\footnote{\url{https://github.com/dendrograms/astrodendro}}, which identifies the changing topology of the surfaces as a function of contour levels and extracts a series of hierarchical structures over a range of spatial scales \citep{Astroden2008}, to identify leaves in the 1.3~mm continuum maps of all fields.  The \astrodendro{} parameters are set to a minimum threshold \texttt{min\_value} of $5\sigma_{\mathrm{cont}}$ for structure identification, a \texttt{min\_delta} of $3\sigma_{\mathrm{cont}}$ to distinguish individual leaves, and a minimum leaf size \texttt{min\_npix} to the number of pixels within one synthesized beam. Among all leaves identified by \astrodendro{}, we then select, through visual inspection, those that show a radially decreasing temperature profile over at least four synthesized beams as the internally heated hot core sample in this work. In total, 83 of the resolved internally heated hot cores were chosen for further analysis and are listed in Table~\ref{tab: hot cores}. One example of the 83 hot cores extracted by \astrodendro{} is outlined in cyan in the lower-left panel of Figure~\ref{fig:image_example}. The remaining core structures are displayed in the complete figure set online.

As reported in \citet{Meng2025}, some hot cores have already evolved to host hyper-compact or ultra-compact H\,\textsc{ii} regions. For these sources, the continuum emission arises from a combination of thermal dust emission and free-free emission from ionized gas. In this work, we refer to them as dust-ff hot cores. To distinguish dust-ff hot cores from the full hot core sample, we first extract the H30$\alpha$ spectra at the peak continuum positions of the hot cores, and then classify a core as a dust-ff hot core if H30$\alpha$ emission is detected at a significance level greater than $3\,\sigma_\mathrm{line}$.
In total, 11 hot cores are identified as dust-ff hot cores. They are marked with an asterisk (*) in Column~(1) of Table~\ref{tab: hot cores}.
It is worth noting that the term dust-ff hot cores used here refers to cores with relatively strong free-free contributions identified through H30$\alpha$ detection. The absence of detected H30$\alpha$ emission in other cores does not necessarily imply the absence of free-free emission.

\subsection{The physical properties of hot molecular cores}
\label{sec: The physical parameters of hot cores}

\subsubsection{Temperature profile}\label{sec: Temperature profile}

Under the assumption of spherical symmetry, the temperature profile of a protostellar envelope can be described by a power-law function \citep[e.g.,][]{Terebey1993,Lu2014,Gieser2021,Gieser2022,Gieser2023,Lin2022,Jeff2024,Motte2025}:
\begin{equation}
T(r)=50\,\mathrm{K} \times\left(\frac{r}{R_{\mathrm{50K}}}\right)^{-q},
\label{eq: T_pow-law}
\end{equation}
where, $R_{\mathrm{50K}}$ is the radius where the temperature reaches 50\,K, and $q$ is the power-law index of the temperature radial profile.
To quantify the temperature profiles of the hot cores, we first divide the temperature maps of the hot molecular cores into a series of concentric annular bins centered on the continuum peak positions. The radial bin width is set to half a beam, denoted as $\delta r$. Accordingly, for each annular bin, the inner and outer radii are defined as $r_i - \delta r/2$ and $r_i + \delta r/2$, respectively.
For each annular bin, we adopt the median temperature as the representative temperature $T_i$ at radius $r_i$. The uncertainty in $T_i$ is estimated using the median absolute deviation, while the radial bin width is taken as the uncertainty in $r_i$.
These representative points are used in the subsequent fitting procedure.
This method is consistent with that adopted in \citet{Gieser2021,Gieser2022,Gieser2023}.

We then fit Equation~\ref{eq: T_pow-law} to the representative points using the Python Orthogonal Distance Regression (ODR) package in \texttt{scipy}. During the fitting process, we exclude the innermost data point, as it significantly deviates from the power-law behavior and lies within an unresolved region (the grey shadow of Figure \ref{fig: profile}). An example fitting result is shown in the upper-left panel of Figure~\ref{fig: profile}. 
The mean and individual temperature profiles of all HMCs are presented in the upper-left panel of Figure~\ref{fig:mean_profile}.
The optimal $R_\mathrm{50K}$ and $q$ for all hot cores are listed at column (7) and (8) of Table~\ref{tab: hot cores}.

The measured $q$ values range from 0.22 to 1.29, with a median value of 0.52. This is slightly steeper than the $q = 0.4$ profile found in the outer parts of envelopes surrounding massive YSOs by \citet{van2000} and $q = 0.36\pm0.02$ inferred in the samples of \citet{Palau2015,Palau2021}. This discrepancy may be due to our higher resolution, which probes the inner regions of the cores.
\citet{Osorio1999} and \citet{van2000} also report a steeper $q$ value near 2,000 au, closer to the protostar. Additionally, other studies have found larger $q$ values in hot molecular cores as well \citep{Beltran2018,Mottram2020,Gieser2021}.
$Log[R_\mathrm{50K}/\mathrm{au}]$ values range from 3.52 to 5.72, with a median value of 4.27 (or 0.25 pc). 
This suggests that the hot cores are producing thermal feedback that heats the surrounding massive clump, at least on sub-pc scales.

We acknowledge the caveat of this method as shown in \citet{Estalella2024}: it is
intrinsically inconsistent to consider that the temperature is constant along the line
of sight while fitting a temperature power-law with radius (i.e., spherically symmetric
function) simultaneously. This effect was quantified in \citet{Estalella2024}, who show that
this method underestimates the temperature power-law index by about 0.15.
\added{
For ease of comparison with previous studies \citep{Gieser2019,Gieser2022,Gieser2023}, we do not deproject or otherwise correct the observed temperature profiles here. Therefore, the fitted values of $R_{\rm 50K}$ and $q$ should be regarded as projected temperature profile parameters measured in the same way as in previous observational works. However, the projection correction is considered in the \radmc{} model in Section~\ref{sec: The embedded protostellar luminosity}.
}

\subsubsection{Density profile}\label{sec: Density profile}

Theoretically, the isothermal Bonnor–Ebert (BE) spheres (in log-log space) \citep{Ebert1955,Bonnor1956} exhibit a nearly flat density profile at small radii from the center. 
At larger radii, the density approximately follows an inverse-square dependence, i.e., $n \propto r^{-2}$. 
Motivated by this behavior, we adopt a modified Plummer-like profile to describe the envelope density structure \citep[e.g.,][]{Dapp2009,Qin2011,Schmiedeke2016,Stutz2018,Tang2018}, i.e.
\begin{equation}
n(r)=\frac{n_c}{\left[1+\left(r / a\right)^2\right]^{p / 2}},
\label{eq:volume density}
\end{equation}
where $n_c$ is the central volume density in cm$^{-3}$, $a$ represents the radius of the inner flat region in au, and $p$ is the density power-law index. When $r\gg a$, $n(r) \sim r^{-p}$. An analytic expression for the column density is derived by integrating the volume density along the line of sight from negative to positive infinity:
\begin{equation}
N(r) = \sqrt{\pi}\, n_c\, a
\left( 1 + \frac{r^2}{a^2} \right)^{\frac{1-p}{2}}
\frac{\Gamma\!\left(\frac{p-1}{2}\right)}{\Gamma\!\left(\frac{p}{2}\right)} ,
\label{eq: N_plummer}
\end{equation}
where $\Gamma(x)$ is the Gamma function. 

As in Section~\ref{sec: Temperature profile}, we continue to use annular bins on the H$_2$ column density map (see the lower-left panel of Figure~\ref{fig:image_example}) to compute the data points used in the fitting procedure. In this case, we adopt a smaller radial bin width of $\delta r = 1/3$ of the synthesized beam to ensure the stability of the fits when a larger number of fitted parameters are involved. 
Similarly, the ODR package is used to optimize fitted parameters. 
The upper-right panel of Figure~\ref{fig: profile} presents one example of the fitting results.
The mean and individual volume and column density profiles are displayed in the lower panels of Figure~\ref{fig:mean_profile}.
The optimal $n_c$, $a$, and $p$ for all hot cores are listed in column (9), (10), and (11) of Table \ref{tab: hot cores}.

\begin{figure*}[ht!]
    \centering
    \includegraphics[width=18cm]{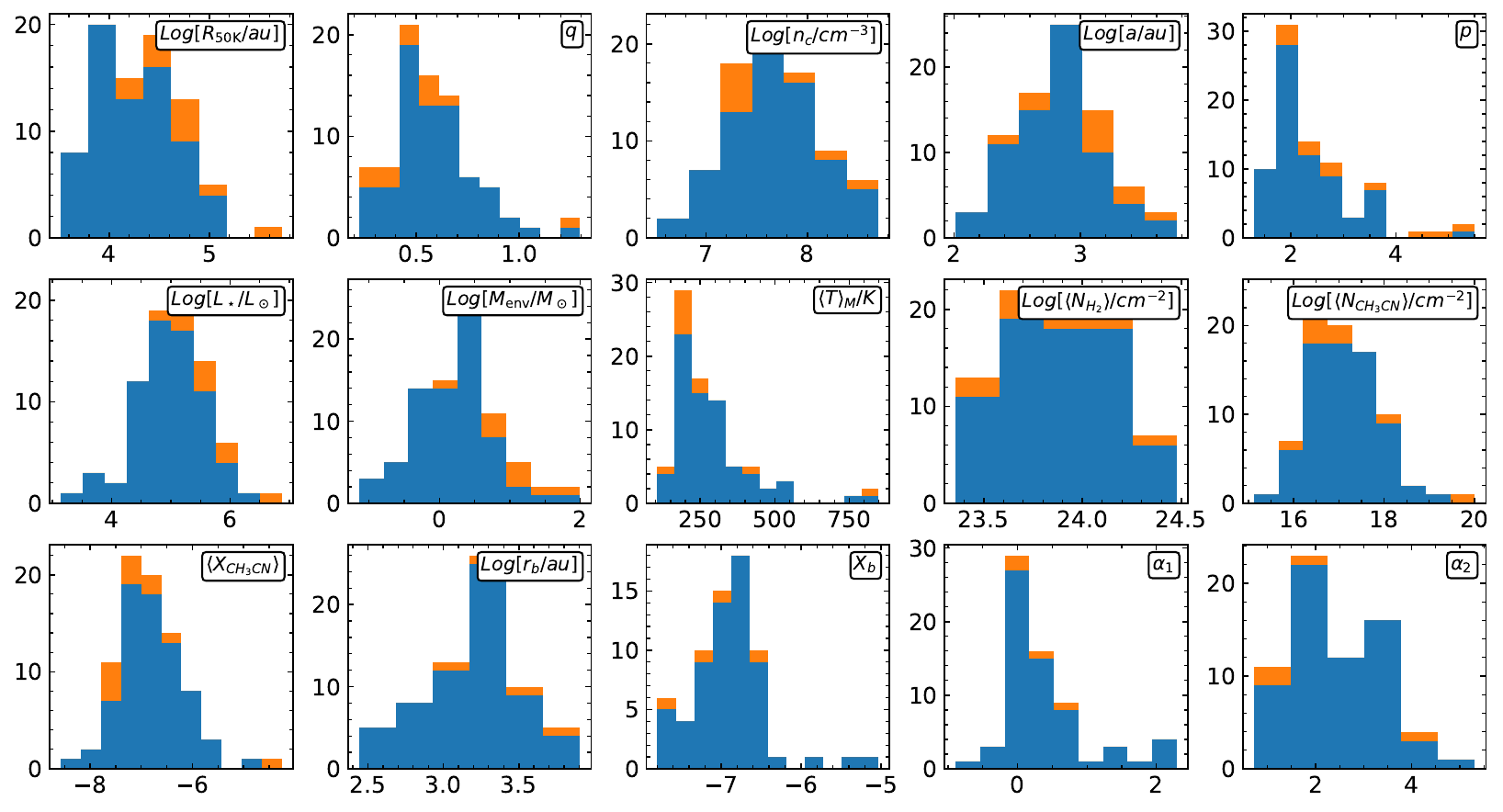}
    \caption{The histograms of the calculated parameters for all hot cores. The orange represents the dust-ff hot cores, and the blue represents the rest of the hot cores.
    }
    \label{fig:hist}
\end{figure*}

The fitted values of $Log[n_c]$, $Log[a]$, and $p$ span ranges of 6.51–8.70 (with a median value of 7.62, or 4.2$\times10^7$ cm$^{-3}$), 2.01–3.75 (with a median value of 2.84, or $\sim$700 au), and 1.29–5.50 (with a median value of 2.10), respectively.
The density profile index $p$ is close to 2 for the majority of cores, consistent with the value expected for an isothermal BE sphere
and with the expected value for
accreting clumps (e.g., \citealt{Gomez2021}).
For $p = 2$, the Jeans mass increases with radius at the same rate as the total enclosed mass, such that the balance between internal pressure and self-gravity can be maintained at all sufficiently large radii in an isothermal BE sphere \citep{2025ifsf.book.....V}.
The histograms of these parameters can be found in Figure~\ref{fig:hist}.

\subsubsection{The embedded protostellar luminosity}\label{sec: The embedded protostellar luminosity}

The luminosity of a protostar is a fundamental parameter for understanding theories of star formation \citep{Myers1998,Offner2011}. Traditionally, astronomers have estimated the bolometric luminosity of clumps by fitting spectral energy distributions (SEDs) spanning from millimeter to infrared wavelengths (e.g., \citealt{Konig2017,Urquhart2018}). With the advent of interferometric observations, core-scale structures can now be well resolved at angular resolutions of $\sim$0.1\arcsec{}--1\arcsec{}. However, the lack of resolution-matched observations in the far-infrared regime makes it difficult to reliably estimate the luminosity of individual protostars through SED fitting alone. 
In our sample, the temperature peak obtained from \ch3cn\,(12--11) coincides with the column density peak, and the temperature decreases with radius, which indicates the hot gas envelope is internally heated by the radiation from the central protostar. Therefore, we can estimate the embedded protostellar luminosity from the observed temperature and density profiles through radiative transfer simulations.
\radmc{} \footnote{\url{https://github.com/dullemond/radmc3d-2.0}} \citep{Dullemond2015}, a publically available Monte-Carlo based radiative transfer code, is suitable to do this.
The workflow illustrated in Figure~\ref{fig:workflow} briefly summarizes how we constrain the protostellar luminosity from the observationally derived density and temperature profiles using \radmc{}. 
A detailed description of the procedure is provided below.

We use the \radmc{} to compute the temperature profiles of a spherical dusty envelope surrounding a central star. 
The observed density profile from Section \ref{sec: Density profile} (here, converted to the dust density) is adopted as the spherical envelope density structure in the simulations, with an inner radius set to one quarter of the synthesized beam and an outer radius of 20{,}000~au. For the dust opacity, we adopt the \citet{Ossenkopf&Henning1994} dust model with thin ice mantles coagulating at a density of $10^{6}\,\mathrm{cm}^{-3}$, commonly referred to as the OH5 model (column~5 of their Table~1). We assume a simple blackbody spectrum for the central source.
No external radiation field was included in the modeling. This is justified by the fact that the regions probed in our analysis correspond to the dense inner parts of the cores, which are expected to be well shielded by the surrounding cloud \citep{Launhardt2013}. As a result, external heating from interstellar radiation field (ISRF) is unlikely to play a dominant role in setting the temperature structure, particularly at the radii constrained by our data. Instead, the temperature profile is expected to be primarily governed by internal heating from the embedded source.
The effective radius of the central source is fixed to one quarter of the synthesized beam, and the corresponding effective temperature is automatically computed according to the Stefan--Boltzmann law. 
The luminosity $L_\star$ is treated as only free parameter and is optimized by minimizing the $$\chi^2 = \sum \limits _{i=1}^{N} \left(\frac{Log[T_\mathrm{obs}^{i}]-Log[T_\mathrm{mod}^i]}{\sigma(Log[T_\mathrm{obs}^i]) }\right)^2$$ value between the model temperature profile $T_\mathrm{mod}^i$ and the observed temperature profile $T_\mathrm{obs}^{i}$ (same as the points used to fit in Section~\ref{sec: Temperature profile}). The optimization is carried out using the \texttt{minimize} function in \texttt{scipy}.
All best-fit $Log[L_\star/L_\odot]$ are listed in column (12) of Table~\ref{tab: hot cores}, and they fall in the range 3.1--6.8 with a median value of 5.1.
\added{
Under the assumption of the ZAMS mass-luminosity relation, the stellar masses fall into the range from 6 to 114 $M_\odot$, which are comparable to the masses of the envelopes.
}
A representative example comparing the model temperature profile obtained with the best-fit luminosity to the observed temperature profile is shown in the lower-left panel of Figure~\ref{fig: profile}.
Note that the luminosity computed using the above method includes the contribution from accretion, internal gravitational contraction and hydrogen fusion.

In the inner regions close to the star, the dust temperature can exceed the sublimation temperature of ice mantles ($\sim$100\,K), such that dust grains may no longer retain their icy mantles. In this case, a bare-grain opacity model, such as OH2 (\citealt{Ossenkopf&Henning1994}, bare grains after coagulation at $10^6$ cm$^{-3}$), may be more appropriate. 
Our current computer program does not implement a mixed opacity model (e.g., adopting OH2 for regions with $T_\mathrm{dust} \gtrsim 100\,\mathrm{K}$ and OH5 for colder regions). 
Instead, we tested the effect of different opacity models by performing \radmc{} calculations using the OH2 opacity for the entire envelope. We find that the resulting luminosity differs by less than 1\% compared to that obtained with the OH5 opacity model.

We examine the relationships between the protostellar luminosities derived using the above method and the bolometric luminosities and masses of the ATLASGAL clumps (Figure~\ref{fig:scatter3}). We find that, for the vast majority of sources, the protostellar luminosities agree with the clump bolometric luminosities to within an order of magnitude and the two luminosities thus exhibit a strong correlation (Pearson coefficient = 0.68). This suggests that the luminosity of a clump is largely dominated by its most luminous embedded massive protostar.
In addition, the protostellar luminosity shows a positive correlation with clump mass (Pearson coefficient = 0.69), indicating that more massive clumps tend to host more luminous protostars. 
This will be discussed in detail in Section~\ref{sec: Coevolution of Clump and Massive Core}.

\added{
As discussed in Section~\ref{sec: Temperature profile}, the observed temperature profiles are measured in projected cylindrical annuli, whereas the \radmc{} model described above, hereafter the original \radmc{} model, produces a temperature profile as a function of spherical radius. Directly comparing these two profiles may therefore introduce a projection-related bias.
To examine this effect, we implement a projection-corrected \radmc{} model in Appendix~\ref{Projection-corrected}.
The additional assumption introduced in this correction is that the projected cylindrical temperature profile can be approximated as the line-of-sight average of the spherical temperature profile, weighted by the \ch3cn{} volume density.
Figure~\ref{fig:projection_model1} shows that the luminosities derived from the original \radmc{} model and from the projection-corrected \radmc{} model are highly consistent. The relative difference in $Log[L_\star]$ is always less than 10\%, with a maximum value of 8.8\%. Therefore, the projection correction introduces only a minor change to the derived logarithmic luminosities and does not affect the subsequent analysis or the conclusions of this work. We therefore keep the luminosities derived from the original \radmc{} model as our fiducial values in the main analysis.
}

\subsubsection{\ch3cn{} abundance profile}

For most of the hot cores, \ch3cn\ abundance profiles $X (r)$ show a broken power-law form. 
Therefore, the following equation is used to fit \ch3cn{} abundance profile:
\begin{equation}
 X(r) =
\begin{cases}
X_b \left( \dfrac{r}{r_b} \right)^{-\alpha_1}, & r < r_b \\
X_b \left( \dfrac{r}{r_b} \right)^{-\alpha_2}, & r \ge r_b
\end{cases} , 
\label{eq:abundance}
\end{equation}
where ($r_b, X_b$) is the break point of the power-law, and $\alpha_1$, $\alpha_2$ are the inner and outer power-law indices, respectively. 
The best fitted $r_b$, $ X_b$, $\alpha_1$, and $\alpha_2$ are [276, 8{,}060]~au with a median value of 1{,}920~au, [-7.8, -5.0] with a median value of -6.9, [-0.89, 2.29] with a median value of 0.17, and [0.72, 5.31] with a median value of 2.19, respectively.
The mean and individual abundance profiles are displayed in the upper-right panel of Figure~\ref{fig:mean_profile}.
The histograms of these parameters are displayed in Figure~\ref{fig:hist}.
An example of the broken power-law profile of \ch3cn{} is displayed in Figure~\ref{fig: profile} (lower-right panel)  and the best-fit paremeters of all cores are listed in Table~\ref{tab: hot cores}.
We will discuss more about \ch3cn{} abundance profile in Section~\ref{sec: ch3cn profile}.

\subsubsection{Mean and median physical parameters}

We compute the mass-averaged temperature $\langle T \rangle_M$ by averaging the temperature map within the \astrodendro{}-identified structures, weighted by column density. 
The derived $\langle T \rangle_M$ values range from 105 to 847\,K, with a median value of 248\,K.
The envelope gas mass $M_\mathrm{env}$ (= $\mu m_\mathrm{H}\int{N_\mathrm{H_2}\mathrm{d}A}$, $\mathrm{d}A$ is the  area element) is also calculated by summing the column density map within the structures identified by \astrodendro{}.
The resulting envelope masses span from 0.07 to $\sim100\,M_\odot$, with a median of $2.1\,M_\odot$.
The H$_2$ column density $\langle N_\mathrm{H_2} \rangle$, \ch3cn{} column density $\langle N_\mathrm{CH_3CN} \rangle$, and \ch3cn{} abundance $\langle X_\mathrm{CH_3CN} \rangle$ are computed as the median values of all pixels in the corresponding maps within the \astrodendro{}-identified structures.
The values of $Log[\langle N_\mathrm{H_2} \rangle]$ range from 23.35 to 24.47, with a median of 23.87, while $Log[\langle N_\mathrm{CH_3CN} \rangle]$ spans 15.12--20.00, with a median of 16.92. 
This median \ch3cn{} column density is close to the value of 17.04 measured toward the high-mass star forming region Sagittarius~B2(N2) \citep{Belloche2016}.
The derived \ch3cn{} abundances $\langle X_\mathrm{CH_3CN} \rangle$ span a wide range, from -8.57 to -4.25, with a median value of -6.94.
Notably, two cores exhibit abundances exceeding -5, which is unusually high and potentially unreliable. 
This apparent enhancement could arise from an underestimation of the H$_2$ column density, or the incorrect \ch3cn{} spectral line fitting due to the optically thick spectra.
All these parameters are listed in Table~\ref{tab: hot cores}.
The histograms of all parameters calculated in Section~\ref{sec: The physical parameters of hot cores} are shown in Figure~\ref{fig:hist}.

\subsection{Correlation Matrix}
\label{sec:matrix}

\begin{figure*}[ht!]
    \centering
    \includegraphics[width=18cm]{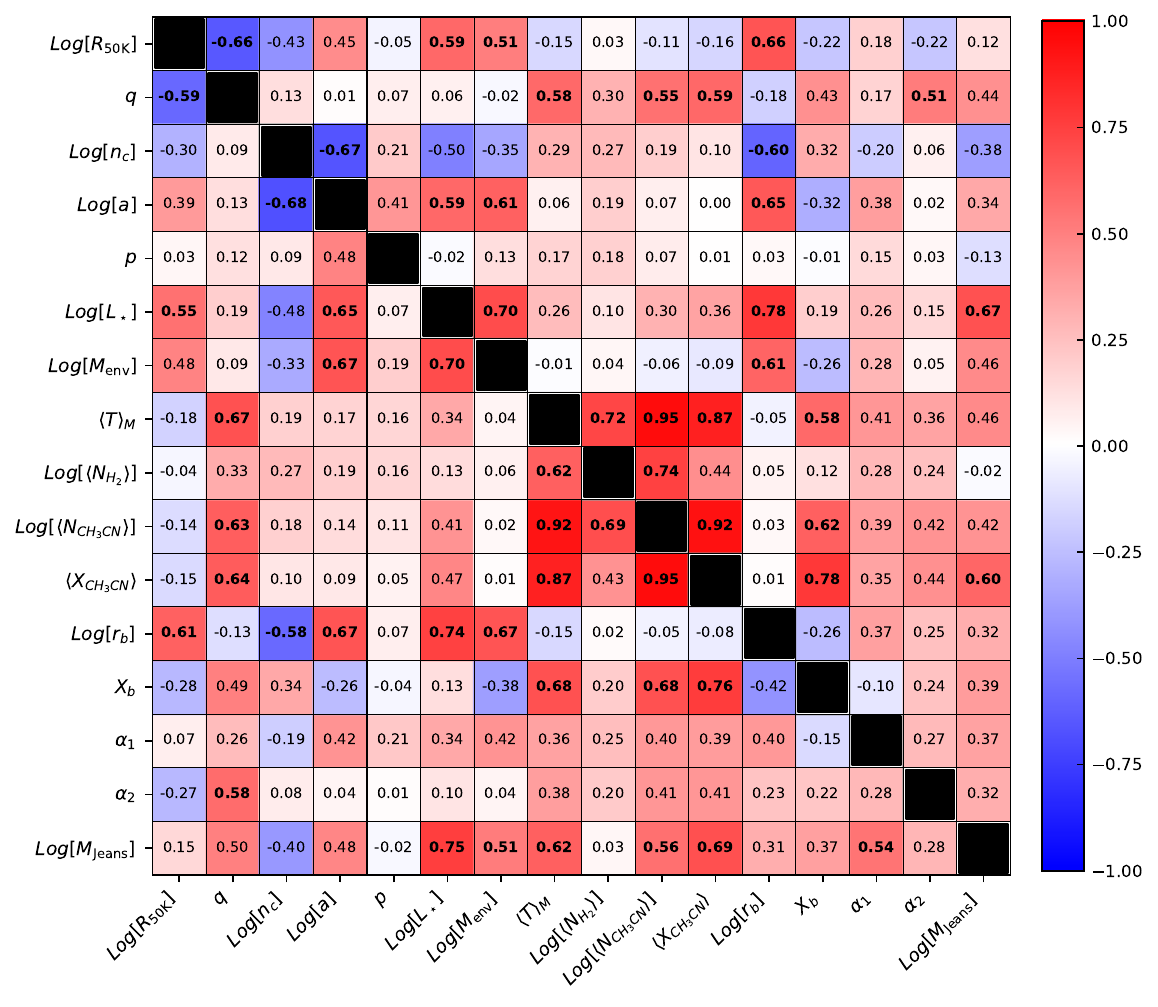}
    \caption{Pearson correlation (lower triangle) and Spearman correlation (upper triangle) matrices for all hot cores. Pearson correlation coefficients are used to quantify linear relationships, while Spearman rank correlation coefficients are adopted to assess monotonic trends that may be non-linear. The bold marks the parameter pairs whose absolute correlation coefficients are above 0.5,  for which linear fitting parameters are listed in Table~\ref{tab:correlation}.
    }
    \label{fig:corr_matrix}
\end{figure*}

\begin{figure*}[ht!]
    \centering
    \includegraphics[width=18cm]{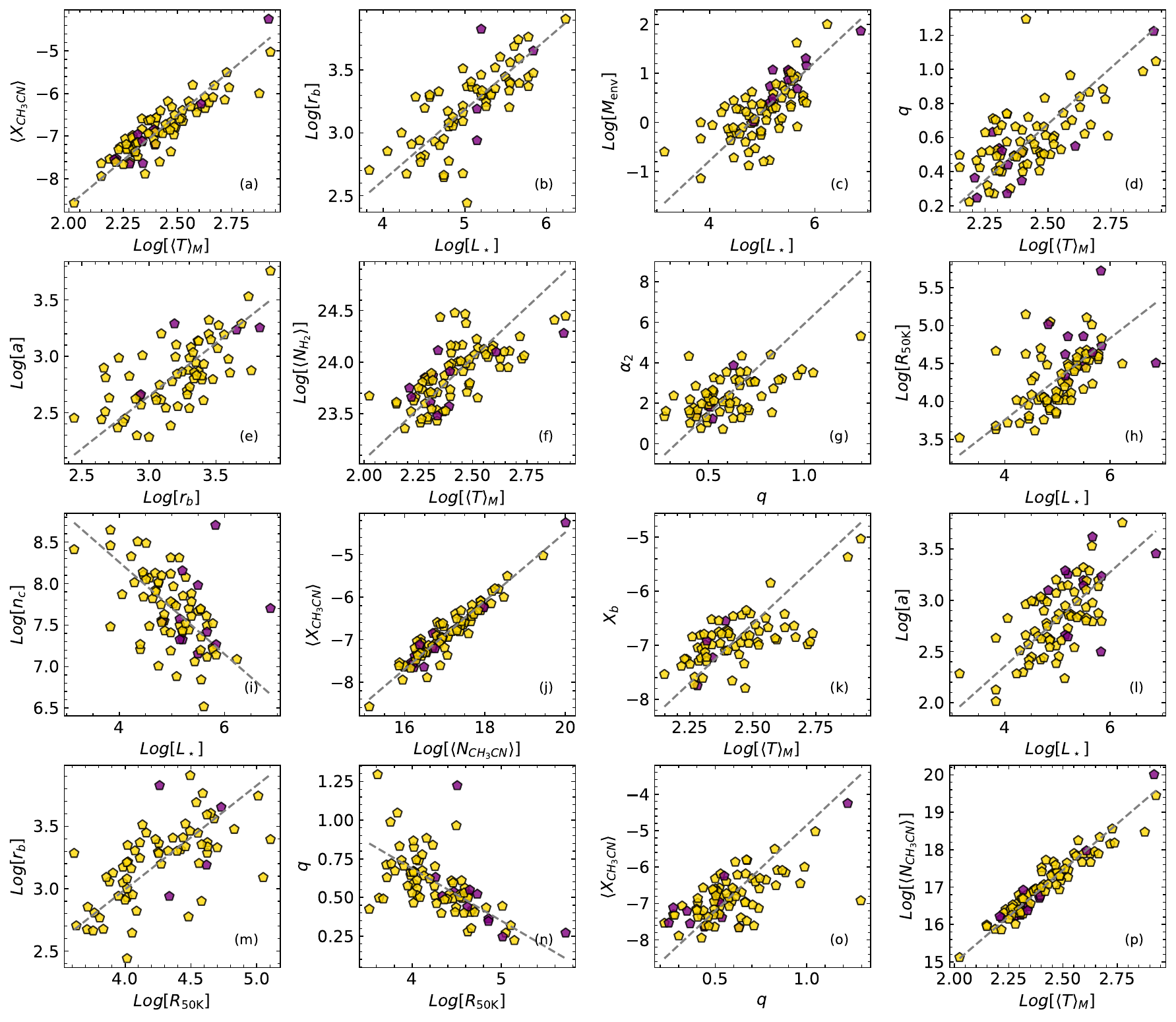}
    \caption{The scatter diagrams between different parameters. The purple pentagons present dust-ff hot cores,  and the gold represent the rest of hot cores. The gray dashed line is the best fitting result, whose slope and intercept can be found in Table~\ref{tab:correlation}.  
    }
    \label{fig:scatter2}
\end{figure*}

\begin{figure*}[ht!]
    \centering
    \includegraphics[width=18cm]{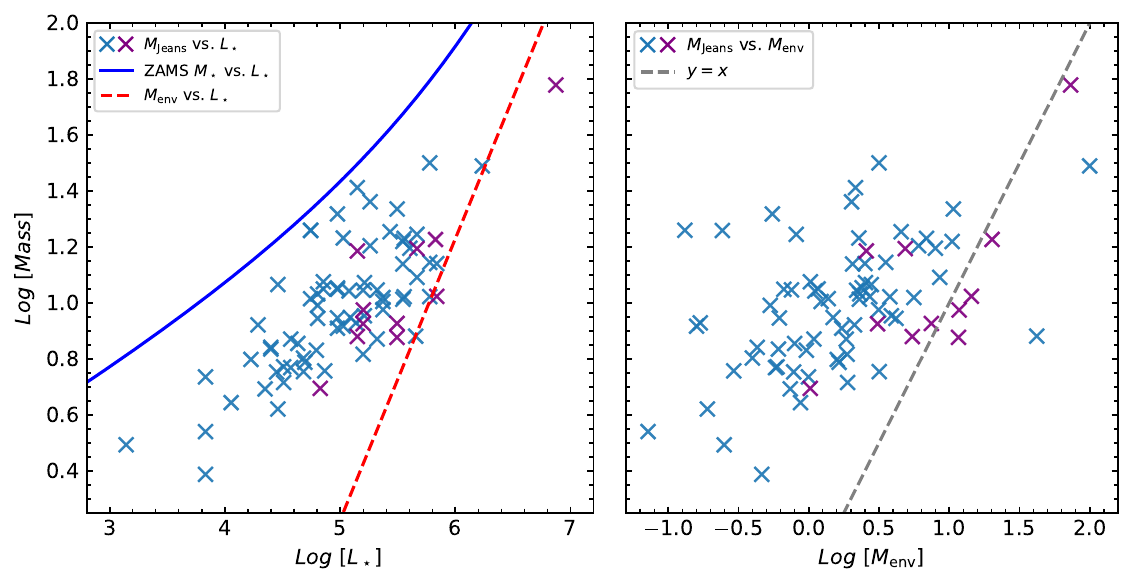}\hfill
    \caption{Left panel: comparison of different mass scales as a function of the embedded protostellar luminosity, $L_\star$. The crosses show the thermal Jeans mass, $M_\mathrm{Jeans}$, versus $L_\star$. The blue solid line shows the ZAMS stellar mass, $M_\star$, as a function of $L_\star$ \citep{2022A&A...658A.125S,Hennebelle2024}. The red dashed line shows the relation between the envelope mass, $M_\mathrm{env}$, and $L_\star$ (Figure~\ref{fig:scatter2} (c)).  Right panel: the scatter between the envelope mass and thermal Jeans mass. The dashed gray line presents $y = x$. The purple crosses present dust-ff hot cores,  and the blue represents the rest of the hot cores.
    }
    \label{fig:scatter1}
\end{figure*}

To systematically examine the relationships among the derived parameters, we construct a matrix of correlation coefficients (Figure~\ref{fig:corr_matrix}). 
For the parameter pairs with the Pearson coefficient above 0.5, linear fits were performed using the ODR method. The best-fit slopes and intercepts are reported in Table~\ref{tab:correlation}.
The scatter plots of some selected parameter pairs are shown in Figure~\ref{fig:scatter2}.
Several relations are discussed below.

The three parameter pairs exhibiting the strongest correlations are
$ Log[\langle N_{\mathrm{CH_3CN}}\rangle]$–$\langle X_{\mathrm{CH_3CN}}\rangle$,
$ Log[\langle N_{\mathrm{CH_3CN}}\rangle]$–$\langle T\rangle_M$, and
$\langle X_{\mathrm{CH_3CN}}\rangle$–$\langle T\rangle_M$.
All three pairs have Pearson correlation coefficients exceeding 0.8. Among them, the $Log[\langle N_{\mathrm{CH_3CN}}\rangle]$–$\langle X_{\mathrm{CH_3CN}}\rangle$ relation is the strongest, with a Pearson coefficient of $r_p = 0.95$ and a $p$-value of $3.43 \times 10^{-42}$.
The scatter plot of $Log[\langle N_{\mathrm{CH_3CN}}\rangle]$ versus $\langle X_{\mathrm{CH_3CN}}\rangle$ is shown in Figure~\ref{fig:scatter2}(j), revealing a remarkably tight linear relationship. 
The linear best-fit relation is $\langle X_{\mathrm{CH_3CN}}\rangle = 0.8Log[\langle N_{\mathrm{CH_3CN}}\rangle] -20.55$.
This suggests that, on spatial scales within $\sim$3,000 au (corresponding to the mean size of our sample), variations in the CH$_3$CN abundance relative to H$_2$ are primarily driven by changes in the CH$_3$CN column density, while the H$_2$ column density exhibits only modest variation, remaining within one order of magnitude (from $10^{23.35}$ to $10^{24.47}$ cm$^{-2}$, see Figure~\ref{fig:hist}).

The scatter plots of the mass-averaged temperature versus the \ch3cn{} abundance and column density are shown in Figure~\ref{fig:scatter2}(a) and (p). Both relations exhibit clear linear trends in log–log space. The corresponding best-fit linear relations are given by $\langle X_{\mathrm{CH_3CN}}\rangle = 4.25Log[\langle T\rangle_M] -17.13$ and $Log[\langle N_{\mathrm{CH_3CN}}\rangle] = 4.96Log[\langle T\rangle_M] +5.06$.
This result contrasts with the abundance analysis of \ch3cn{} in the Sagittarius B2 region reported by \citet{Pols2018}, who found an anti-correlation between abundance and temperature. They suggested that this may be related to stronger feedback processes that destroy \ch3cn{}. 
In contrast to their findings, our positive correlation between abundance and temperature does not show any clear evidence of feedback mechanisms destroying \ch3cn{}. 
Instead, our results are more in line with the predictions of the model by \citet{Rodgers2001}, which simulates a significant enhancement in \ch3cn{} abundance at 300 K compared to 100 K.
More recently, the more comprehensive chemical models developed by \citet{Garrod2022} likewise show that the abundance of \ch3cn{} increases nearly monotonically with temperature during the warm-up phase, further supporting a strong link between \ch3cn{} abundance and high-temperature chemical evolution.


The $L_\star$–$M_\mathrm{env}$ ($r_p = 0.70$, $r_s = 0.70$) scatter plot is shown in Figure~\ref{fig:scatter2}(c). A similar mass-luminosity diagram is presented in \citet{Urquhart2018} and \citet{Motte2025}. In \citet{Urquhart2018}, the massive star-forming (MSF) clump mass-luminosity relation is fitted as $ Log[M_\mathrm{clump}] = 0.76 Log[L_\mathrm{clump}] - 0.07 $.
However, on the core scale, we obtain the best-fit linear relation as $ Log[M_\mathrm{env}] = 1.01 Log[L_\star] - 4.80 $. 
This empirical relation may provide useful observational constraints for future studies on physical (or theoretical) models of protostellar objects.

Figure~\ref{fig:scatter2}(d) shows that as the mass-averaged temperature increases, the temperature profile index $q$ also increases. A similar correlation is observed in \citet{Gieser2023}.
Figure~\ref{fig:scatter2}(n) indicates an anti-correlation between the temperature profile index and the radius at 50 K ($R_\mathrm{50K}$). The histogram of $R_\mathrm{50K}$ (Figure~\ref{fig:hist}) shows that dust-ff hot cores tend to have larger $R_\mathrm{50K}$.
These results suggest that a more evolved core does not necessarily have a higher temperature profile index, contrary to what is suggested by \citet{Gieser2023}.

Regarding the parameters of the density profile, we find that the central density of the core, $n_c$, is negatively correlated with the protostellar luminosity ($r_p = 0.48$, $r_s = 0.50$, $ Log[n_c] = -0.55 Log[L_\star] +10.47 $, Figure~\ref{fig:scatter2} (i)). This trend may indicate that stronger radiative feedback leads to photo-evaporation of dust in the innermost regions.
In addition, the flat radius of the density profile is positively correlated with protostellar luminosity ($r_p = 0.65$, $r_s = 0.59$, $Log[a] = 0.46\,Log[L_\star] + 0.52$, Figure~\ref{fig:scatter2}(l)), following a power-law relation of $a \propto L_\star^{0.46}$. This exponent is comparable to that of the dust destruction radius as a function of luminosity $r_d \propto L_\star^{0.5}$ \citep{Krumholz2015,Krumholz2018}.
We did not find a strong correlation (i.e., $r_p > 0.5$) between the density profile index $p$ and any other parameters (see Figure~\ref{fig:corr_matrix}).

\section{DISCUSSION}

\subsection{Implications from \ch3cn{} abundance profile}
\label{sec: ch3cn profile}

From our measurements of the \ch3cn{} abundance profiles within hot core envelopes, we find that the vast majority of sources exhibit radially decreasing abundances away from the continuum peak. From the innermost regions, where the \ch3cn{} abundance is highest, to the outer envelope, the abundance typically decreases by one to two orders of magnitude (an example is shown in the lower-right panel of Figure~\ref{fig: profile}) and follows a broken power-law beyond $\sim$1{,}000 au. We ran a simple chemical model with a gas-grain reaction network to study the \ch3cn{} abundance profiles using the \texttt{GGCHEMPY} code \footnote{\url{https://github.com/JixingGE/GGCHEMPY}} \citep{Ge2022}. The details of the model can be found in Appendix \ref{sec:A simple chemical model of hot core}. Models at early evolutionary stages of $<10^4$ years can well reproduce the \ch3cn{} abundances at larger radii ($r > 1{,}000$ AU) but fail at small radii ($r < 1{,}000$ AU). This simple model indicates that the breaking radius ($r_b$) of the \ch3cn{} abundance profile likely marks the boundary between the inner hot region ($> 200$ K) and the outer warm region ($< 120$ K). Within the hot core’s inner region, the gas-grain chemical model underestimates the observed \ce{CH3CN} abundance and may indicate that \ce{CH3CN} in the inner region is primarily formed via other mechanisms such as efficient gas-phase ion-molecule reactions following the destruction of refractory organics (e.g., \ce{CH3+ + HCN -> CH3CNH+ -> CH3CN}, \citealt{Nazari+2023}). In the outer warm region, its abundance is dominated by desorption of ice-mantle precursors and influenced by FUV-driven photochemistry. With lower density and higher FUV photons that refers to the ISRF, \ce{CH3CN} abundance decreases toward the outer region. 

Recently, a new and more comprehensive astro-chemical modeling study of hot molecular cores demonstrated that increasing the warm up timescale can significantly enhance the gas-phase abundance of \ch3cn{} \citep{Garrod2022}. In that work, three chemical models with different warm up timescales were considered, reaching 200~K over $5 \times 10^{4}$~yr (fast), $2 \times 10^{5}$~yr (medium), and $1 \times 10^{6}$~yr (slow). The slow warm up model produces a final \ch3cn{} abundance that is higher by approximately 1.6 orders of magnitude compared to the fast model, comparable to the abundance contrast we observe between the inner and outer regions of the envelopes.
They further suggested that differences in the abundances of N-bearing species, such as \ch3cn{}, among different sources are primarily related to the amount of time available for gas-phase chemistry at high temperatures ($>100$~K). 
Following this interpretation, the large \ch3cn{} abundance contrasts observed between the central and outer envelope regions in our sample may reflect differences in the duration of the high-temperature phase. This scenario is naturally consistent with internal heating, in which the central regions of the envelope reach high temperatures earlier and therefore experience a longer period of efficient gas-phase chemistry.
Within this framework, future studies may exploit radial \ch3cn{} abundance profiles as a potential diagnostic of the timescales over which different envelope radii are heated to high temperatures, thereby providing new insights into the dynamical and thermal evolution of protostellar envelope heating.

In addition, a small fraction of hot cores show a negative value for the first \ch3cn{} abundance index $\alpha_1$ (see Figure~\ref{fig:hist}), implying a slight decrease in abundance toward the immediate vicinity of the protostar. While the underlying cause remains uncertain, this trend may be indicative of partial \ch3cn{} destruction in the inner regions, potentially associated with photo-dissociation processes induced by protostellar radiation.
Interestingly, the protostellar luminosity $L_\star$ exhibits a strong correlation ($r_p = 0.74$, $r_s = 0.78$, Table~\ref{tab:correlation}) with the break radius $r_b$ of the CH$_3$CN abundance profile, providing further evidence for a feedback mechanism in which the protostar influences the chemical abundance distribution in the surrounding envelope.
The scatter of $L_\star$ versus $r_b$ is shown in Figure~\ref{fig:scatter2}(b), and the best-fit line is $Log[r_b] = 0.56Log[L_\star]+0.37$.

\subsection{The effect of thermal feedback on core fragmentation}
\label{sec: fragmentation}

Numerical simulations indicate that perturbed molecular cloud cores with several initial Jeans masses undergo efficient fragmentation into binary systems. This fragmentation is subsequently suppressed by thermal heating as the collapsing clouds become optically thick \citep{Boss1983,Boss1984,Boss1986}.
In addition, simulations that incorporate thermal (radiative) feedback, in contrast to those without such heating, substantially reduce the overproduction of brown dwarfs and enhance the formation of massive stars \citep[e.g.,][]{Krumholz2006,Krumholz2007,Krumholz2008,Bate2009,Bate2012,Krumholz2010,Krumholz2012,Offner2009,Urban2010}.
While numerical simulations suggest an important role is played by thermal feedback, observational constraints are still limited. In particular, whether thermal feedback can effectively inhibit further subsequent fragmentation within dense cores, thereby facilitating the formation of massive stars, remains an open question. 
Therefore, in this section, we use our sample, for which the envelope temperature fields are spatially resolved, to investigate the extent to which thermal feedback influences subsequent fragmentation within dense cores.

Protostellar luminosity traces the total radiative energy released by the central source and serves as a direct proxy for thermal feedback imparted to the envelope.
In addition, the classical thermal Jeans mass is commonly used to characterize fragmentation criteria in structures: thermally supported structures possess critical mass and length scales, above which they become gravitationally unstable and undergo fragmentation \citep{Beuther2025}.
We examine the scatter distribution between the luminosity of the internal protostar (although derived from the surrounding envelope) and the Jeans mass of the envelope (see Figure~\ref{fig:scatter1} left panel). The thermal Jeans masses are calculated using the following equation \citep{Wang2014, Liu2017}:
\begin{equation}
M_{\mathrm{Jeans}}=0.877 M_{\odot}\left(\frac{T}{10 \mathrm{~K}}\right)^{3 / 2}\left(\frac{n}{10^5 \mathrm{~cm}^{-3}}\right)^{-1 / 2} ,
\end{equation}
where $T$ is replaced with the mass-averaged temperature $\langle T \rangle _M$ and $n$ is replaced with the core-averaged volume density $\langle n\rangle$($=\int n(r) \mathrm{d}V/(4/3\pi R^{3})$, where $\mathrm{d}V$ is the volume element).
The calculated Jeans mass $M_{\mathrm{Jeans}}$ ranges from 2.4 to 60.0 $M_\odot$ with a median value of 10.0 $M_\odot$.
The scatter in log–log space reveals a strong correlation between the Jeans mass and the protostellar luminosity. The Pearson and Spearman correlation coefficients are 0.75 and 0.67 (see Figure~\ref{fig:corr_matrix}), respectively.
Furthermore, for the majority of sources in our sample, the Jeans masses enhanced by thermal feedback exceed the observed envelope masses (Figure~\ref{fig:scatter1} right panel).
These results suggest that, from an observational perspective, thermal feedback can effectively increase the Jeans mass of the envelope and thereby is slowing down their fragmentation or even completely inhibiting it, in agreement with the numerical simulations \citep{Offner2009,Krumholz2007,Krumholz2011,Myers2013}.
Our findings are further supported by several observational studies.
\citet{Tang2022} explored the mass and separation distribution of fragments in the feedback-dominated W51-North cloud, and found that they are consistent with a Jeans fragmentation process operating at high temperatures in the range 200-400 K. 
Similarly, \citet{Palau2024} reported a deficit of low-mass proto-brown dwarfs in
the nearby molecular clouds with higher average temperatures due to feedback from nearby OB associations.
In addition, \citet{Bouy2026} found a deficit of binaries in the hotter cloud.

We fit the scatter between the thermal Jeans mass and the protostellar luminosity with a power-law relation, i.e., $Log[M_\mathrm{Jeans}] = 0.32Log[L_\star]-0.63$. 
The best-fitting trend can equivalently be expressed as $L_\star \propto {M_\mathrm{Jeans}}^{3.1\pm0.3} $, whose index is close to (slightly shallower than) the mass–luminosity relation $L_\star \propto {M_\star}^{3.5} $ for main-sequence stars \citep{Salaris2006}. 
This similarity may suggest that the local Jeans mass plays an important role in setting the final stellar mass. However, this interpretation remains speculative and requires further investigation in the future.

\subsection{Coevolution of natal clumps and the most massive protostars}
\label{sec: Coevolution of Clump and Massive Core}

\begin{figure*}[ht!]
    \centering
    \includegraphics[width=18cm]{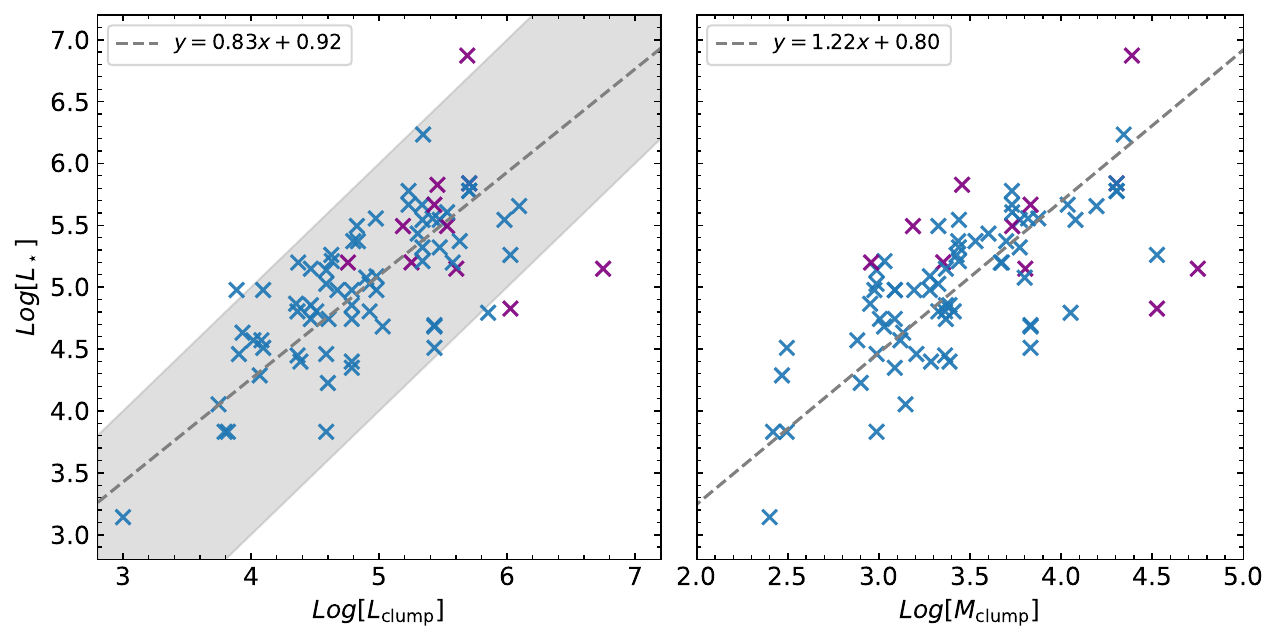}
    \caption{The protostellar luminosity versus the clump luminosity (left panel) and mass (right panel).  The grey shadow shows the region between $y=x\pm1$. The purple crosses present dust-ff hot cores,  and the blue represents the rest of the hot cores. The gray dashed line is the best-fitting result.
    }
    \label{fig:scatter3}
\end{figure*}

\added{
Before proceeding with the discussion, we note that some of the interpretations below should be treated with caution because they may be partly affected by statistical effects, as discussed in Section~\ref{sec: The improvements needed in the future}.
}

Numerous studies have shown that massive clumps and their embedded massive cores are connected through filamentary gas accretion flows, which transport material from clump scales down to core scales \citep[e.g.,][]{Peretto2013,Peretto2014,Lu2018,Olguin2023,Xu2023,Yang2023,Zhou2024,Mai2025,Olguin2025}. In recent years, ALMA observations have revealed a positive correlation between the clump mass and the mass of its most massive core (MMC), providing strong evidence for the coevolution of massive clumps and their MMCs \citep{Anderson2021,Xu2024,Elia2026}.
\citet{Xu2024} argue that such multiscale gas accretion flows are the primary driver of this coevolution and propose a dynamic picture of the clump–core connection: initially, Jeans fragmentation produces a population of dense cores whose masses are not linked to the clump-scale gravitational potential or turbulence; as the clump evolves, continuous gas accretion across multiple scales progressively builds a connection between clump and core properties, leading to the observed mass correlation.

However, this dynamic picture does not explicitly consider the impact of thermal feedback from massive protostars. 
In this section, we investigate how protostellar thermal feedback may modify the clump–core coevolution scenario.
As established in previous studies, more massive clumps tend to host more massive cores \citep{Anderson2021,Xu2024}. 
However, the mass of a core does not directly represent the protostellar mass, but rather the gas mass of the protostellar envelope. 
A key question in this dynamic process is 
how much of the gravitational energy released during accretion is radiated as luminosity that heats the surrounding gas.
Considering an extreme case where only a small fraction of the accreted mass is contributed into radiative output (weaker thermal feedback), the continuous inflow of material from larger scales would increase the core mass while keeping the gas temperature comparatively low. 
As a result, the instantaneous core mass would systematically exceed the local Jeans mass, favoring fragmentation rather than continued accretion.
Therefore, in the absence of effective thermal feedback, the mere availability of a large mass reservoir in the clump would not be sufficient to form more massive cores. 
In contrast, if the embedded protostar is able to radiate sufficient luminosity to heat its envelope, the local Jeans mass would increase accordingly. This thermal feedback can effectively suppress fragmentation, thereby promoting the sustained growth of massive cores.
It is worth noting that additional mechanisms, such as turbulence, may still lead to the formation of more massive cores \citep{Xu2023b}, which is beyond the scope of this work.
We find a strong positive correlation between massive protostellar luminosity and clump mass, with the protostellar luminosity remaining comparable in magnitude to the clump bolometric luminosity (Figure~\ref{fig:scatter3}). 
The Pearson and Spearman correlation of $Log[L_\star]-Log[M_\mathrm{clump}]$ coefficients are 0.68 and 0.66, respectively (see Figure~\ref{fig:corr_matrix}). Both correlations are statistically significant, with p-values below 0.01.
This may suggest that massive protostars forming in more massive clumps tend to reach higher luminosities (A similar result can be found in \citealt{Dib2023}). The enhanced radiative output increases the local Jeans mass, thereby facilitating continued core accretion and naturally giving rise to the observed correlation between core mass and clump mass reported in \citet{Anderson2021,Xu2024}.

Furthermore, a linear fit to the $Log[M_\mathrm{clump}]$–$Log [L_\star]$ scatter yields a relation of $L_\star \propto {M_\mathrm{clump}}^{1.22}$.
Assuming a stellar mass–luminosity relation of $L_\star \propto {M_\star}^{\beta}$, this implies $M_\star \propto {M_\mathrm{clump}}^{1.22/\beta}$.
Similarly, the best-fitting relation for the $M_\mathrm{env}$–$L_\star$ is $L_\star \propto {M_\mathrm{env}}^{1.01}$, corresponding to $M_\star \propto {M_\mathrm{env}}^{1.01/\beta}$.
The combined relations, $M_\star \propto {M_\mathrm{env}}^{1.01/\beta} \propto {M_\mathrm{clump}}^{1.22/\beta}$, suggest a coevolutionary connection among massive protostars, their natal cores, and the parent clumps. In particular, the protostellar mass appears to be more sensitive to the clump-scale mass reservoir than to the instantaneous mass of the core.
In star clusters the mass of the most massive star is also found to depend on the total stellar mass of the cluster, following $M_{\star \mathrm{max}} \propto {M_{\star \mathrm{cluster}}}^{0.58}$ \citep{Weidner2013}. This scaling is comparable to our derived $M_\star$–$M_\mathrm{clump}$ relation when adopting $\beta \simeq 2.1$. 
This similarity suggests that the $M_{\star\mathrm{max}}$–$M_{\star\mathrm{cluster}}$ relation may already be imprinted during the protocluster phase.

\subsection{Caveats}
\label{sec: The improvements needed in the future}


Our sample spans a wide range of distances (1.3\,kpc -- 12.9\,kpc); therefore, potential distance-related biases need to be considered. 
We first restrict our sample to cores within 5~kpc to construct a distance-limited subsample, which contains a total of 56 cores. We then recompute the Pearson correlation coefficients for the parameter pairs that exhibit relatively strong correlations in Figure~\ref{fig:corr_matrix} (listed in Table~\ref{tab:correlation}). The resulting Pearson coefficients ($r_p^\mathrm{lim}$) and corresponding logarithmic p-values ($Log[p_p^\mathrm{lim}]$) for the distance-limited sample are also presented in the column (11) and (12) of Table~\ref{tab:correlation}.
We find that most of the correlations discussed in Section~\ref{sec:matrix} remain strong ($r_p^\mathrm{lim} \geq 0.5$, $p_p^\mathrm{lim}\ll0.05$) within the distance-limited sample, such as $ Log[\langle N_{\mathrm{CH_3CN}}\rangle]$–$\langle X_{\mathrm{CH_3CN}}\rangle$,
$ Log[\langle N_{\mathrm{CH_3CN}}\rangle]$–$\langle T\rangle_M$, and
$\langle X_{\mathrm{CH_3CN}}\rangle$–$\langle T\rangle_M$. The only notable exception is the $Log[L_\star]$--$Log[n_c]$ relation, for which the Pearson correlation becomes significantly weaker ($r_p^\mathrm{lim}=-0.38$, $p_p^\mathrm{lim}=0.003$). This suggests that the majority of the correlations reported in this work are not strongly affected by distance-related biases.
In addition, 
in Appendix~\ref{Possible Distance-related Bias} we present the impact of distance on the observational data and the physical quantities derived in Section~\ref{sec: The physical parameters of hot cores}.
In Figure~\ref{fig:distance1}, several physical quantities exhibit relatively stronger correlations (Pearson coefficient $r_p > 0.5$)  with distance. Consequently, some of the correlations presented in Section~\ref{sec:matrix} may be partially influenced by these distance-dependent effects and should therefore be interpreted with caution.
In particular, the last panel of Figure~\ref{fig:distance1} show that the core sizes derived by \astrodendro{} increase with distance. This trend may indicate that sources observed at different distances do not necessarily represent the same type of physical object: nearby sources may correspond to individual protostars, whereas more distant sources may contain a group of protostars. However, the strong correlations in Figure \ref{fig:scatter2} exhibited by sources at different distances may indicate that the feedback effect of a single star is highly similar to the integrated feedback effect of multiple stars.

The measurement of the dust temperature profile is crucial for studies of star formation, yet it remains notoriously difficult to constrain observationally.
In deriving the dust temperature profiles, we first estimate the gas kinetic temperature using the rotational temperature obtained from \ch3cn\,(12--11). The dust temperature is then assumed to be equal to the gas kinetic temperature.
Two aspects of this approach can be improved in future studies. First, although \ch3cn\,(12--11) traces relatively warm and inner regions compared to lower-excitation tracers such as \ch3cn\,(5--4) used in \citet{Gieser2023} and H$_2$CO used in \citet{Gieser2021}, it is still not capable of probing the innermost and hottest regions of HMCs. Therefore, the rotational transitions in vibrationally excited states will be required in future studies to better constrain the gas temperature structure in these regions.
Second, although theoretical models suggest that gas and dust are thermally coupled at sufficiently high densities ($n \ge 10^{5} \mathrm{cm^{-3}}$) where $T_\mathrm{kin}$ can approach $T_\mathrm{dust}$ within $\sim$10--20\% \citep{Goldsmith2001,Young2004,Ivlev2019}, The complex organic molecules (COMs) may be located in regions with different temperature from the regions that dominate the dust distribution. 
This was already noted by \citet{Motte2025} in the ALMA-IMF sample, who reported that the dust temperatures derived from SED fitting are systematically lower than the rotational temperatures inferred from COM lines. 
If a similar discrepancy exists in our sample, the protostellar luminosities derived in Section~\ref{sec: The embedded protostellar luminosity} may be overestimated. 
Therefore, a more reliable approach would be to derive the dust temperature profile directly from multi-band SED fitting rather than adopting the gas temperature as a proxy. 
For example, \citet{2024A&A...687A.217D} employed the point process mapping (PPMAP) technique to perform SED fitting of far-infrared and submillimeter observations, enabling the construction of dust temperature maps. 
However, the method relies on a Bayesian framework that combines data spanning a wide range of angular resolutions, requiring extrapolation across very different spatial scales. 
As a result, the derived dust temperature structure may be model dependent and subject to systematic uncertainties. 
Multi-band observations with matched angular resolution will therefore be required in future studies to more reliably determine the dust temperature structure.

The calculation in Section~\ref{sec: The physical parameters of hot cores} assumes spherical symmetry when deriving the temperature and density profiles and when modeling the protostellar luminosities. While this assumption provides a reasonable first-order approximation, massive star-forming cores are often characterized by complex internal structures, including accretion disks, bipolar outflow cavities, and asymmetric or clumpy density distributions. These structures can introduce significant deviations from spherical symmetry and may influence the derived temperature and density profiles.
As a result, the simplified geometry adopted in this work may lead to systematic uncertainties in the inferred physical parameters. High angular resolution observations combined with more sophisticated three dimensional radiative transfer modeling will be required in future studies to better capture the intrinsic complexity of massive star-forming environments.

\added{
Finally, the positive $L_\star$--$M_{\rm clump}$ correlation should be interpreted with caution. As discussed in studies of the maximum-stellar-mass--cluster-mass ($m_{\rm max}$--$M_{\rm ecl}$) relation, such trends can be influenced by statistical IMF sampling \citep[e.g.,][]{Weidner2013,Weidner2014}. However, previous Monte Carlo studies have also argued that the observed $m_{\rm max}$--$M_{\rm ecl}$ relation is not fully explained by random sampling alone and may contain physical information about clustered massive-star formation \citep{Weidner2013}. Therefore, the $L_\star$--$M_{\rm clump}$ relation in our sample may include both statistical and physical contributions.
}

\section{conclusions}

In this work, we selected 58 fields from the QUARKS survey, in which the \ch3cn{} emission is well spatially resolved (with emission sizes larger than four times the synthesized beam). This allows us to resolve the physical and chemical structures of hot molecular cores (HMCs).
We summarize our main results as follows:
\begin{enumerate}
    \item Using the multi-transition \ch3cn\,(12--11) emission, we derived temperature and \ch3cn{} column density maps for all 58 fields with \spectuner{}. Combined with the temperature and 1.3~mm continuum maps, we identify 83 core structures whose envelopes exhibit radially decreasing temperature profiles. These cores are internally heated by embedded massive protostars and constitute a suitable sample for investigating thermal feedback in HMCs.
    \item Based on the continuum and temperature maps, we derived H$_2$ column density maps and further constructed radial density profiles for the 83 cores. By combining the observed density and temperature profiles, we performed thermal Monte Carlo radiative transfer simulations to estimate the luminosities of the embedded massive protostars. The derived protostellar luminosities span $Log[L_\star/L_\odot]=3.1$--6.8, with a median value of 5.1, which is comparable to the bolometric luminosities of their natal clumps.
    \item Using the H$_2$ and \ch3cn{} column density maps, we obtained the \ch3cn{} abundance distributions within the envelopes of these HMCs. We find that most cores exhibit radially decreasing abundance profiles, which can be well described by a broken power-law function.
    \item A correlation matrix of all physical parameters was constructed. We find the massive protostellar luminosity is positively correlated with the flat radius $ (Log[a] = 0.46 Log[L_\star] +0.52)$ and mass $ (Log[M_\mathrm{env}] = 1.01 Log[L_\star] - 4.80) $ of the hot molecular core, but negatively correlated with the number density $ (Log[n_c] = -0.55 Log[L_\star] +10.47) $. 
    Here, $a$ and $n_c$ denote the flat radius and central density, respectively, as defined by the Plummer-like parametrization of the density profile.
    \item We find a strong positive correlation between massive protostellar luminosity and the local thermal Jeans mass, with the local Jeans masses systematically exceeding the envelope masses. From an observational perspective, this supports the scenario that thermal feedback from massive protostars can effectively suppress further fragmentation within core structures.
    \item A significant positive correlation is also found between the massive protostellar luminosity and the natal clump mass. This suggests that thermal feedback may play an important role in the coevolution of cores and their parent clumps, providing an extension to previous studies.
\end{enumerate}

\begin{acknowledgments}
\added{We thank the referee for providing constructive comments
that substantially improved the quality of the paper.}
This work was main funded by the National Key R\&D Program of China
under grant Nos. 2022YFA1603100 and 2023YFA1608002, National Science and Technology Major Project 2024ZD1100601, and
Tianshan Talent Training Program 2024TSYCTD0013. 
It was also partially funded by the NSFC under grants 12173075,
12373029 and 12403033, the CAS Light of West China
Program XBZG-ZDSYS-202212, the Tianshan Talent Program
of Xinjiang Uygur Autonomous Region under grant No.
2022TSYCLJ0005, and the Youth Innovation Promotion
Association CAS. 
T.L. acknowledges the support by the National Natural Science Foundation of China (NSFC) through
grants No. 12073061 and No. 12122307, the PIFI program of
the Chinese Academy of Sciences through grant No.
2025PG0009, and the Tianchi Talent Program of Xinjiang
Uygur Autonomous Region.
This work was supported by the Young Scientists Fund of the National Natural Science Foundation of China (Grant No. 12403030).
\added{
J.X.G. thanks the Xinjiang Tianchi Talent Program (2024) and 
the Natural Science Foundation of Xinjiang Uygur Autonomous Region, No. 2025D01D49.}
A.P. acknowledges financial support from the UNAM-PAPIIT IN120226 grant, and the Sistema Nacional de Investigadores of SECIHTI, M\'exico.
G.G. gratefully acknowledges support by the ANID BASAL project FB210003.
This work was performed in part at the Jet Propulsion Laboratory, California Institute of Technology, under contract with the National Aeronautics and Space Administration (80NM0018D0004).
SRD acknowledges support from the Fondecyt Postdoctoral fellowship (project code 3220162) and ANID BASAL project FB210003.
L.B. gratefully aknowledges support from ANID Basal Project FB210003.
AS gratefully acknowledges support by the Fondecyt Regular (project
code 1220610), and ANID BASAL project FB210003. 
AS is gratefully supported by the China-Chile Joint Research
Fund (CCJRF No. 2312). CCJRF is provided by Chinese Academy of Sciences South America Center for Astronomy (CASSACA) and established by
National Astronomical Observatories, Chinese Academy of Sciences (NAOC)
and Chilean Astronomy Society (SOCHIAS) to support China-Chile collaborations in astronomy.
C.W.L is supported by the Korea Astronomy and
Space Science Institute grant funded by the Korea government (MSIT; project No. 2025-1-841-02).

\end{acknowledgments}

\facilities{ALMA}
\software{ASTROPY \citep{2013A&A...558A..33A,2018AJ....156..123A,2022ApJ...935..167A}, CASA \citep{2022PASP..134k4501C},
CARTA \citep{Carta2021},
ASTRODENDRO \citep{Astroden2008},
SPECTUNER \citep{Qiu2025,Qiu2026},
RADMC-3D \citep{Dullemond2015},
GGCHEMPY \citep{Ge2022}
}

\appendix

\renewcommand\thefigure{\Alph{section}\arabic{figure}}
\renewcommand\thetable{\Alph{section}\arabic{table}}

\added{
\section{Projection-corrected \radmc{} Model}
\label{Projection-corrected}
\restartappendixnumbering
}
\begin{figure}[ht!]
    \centering
    \includegraphics[width=18cm]{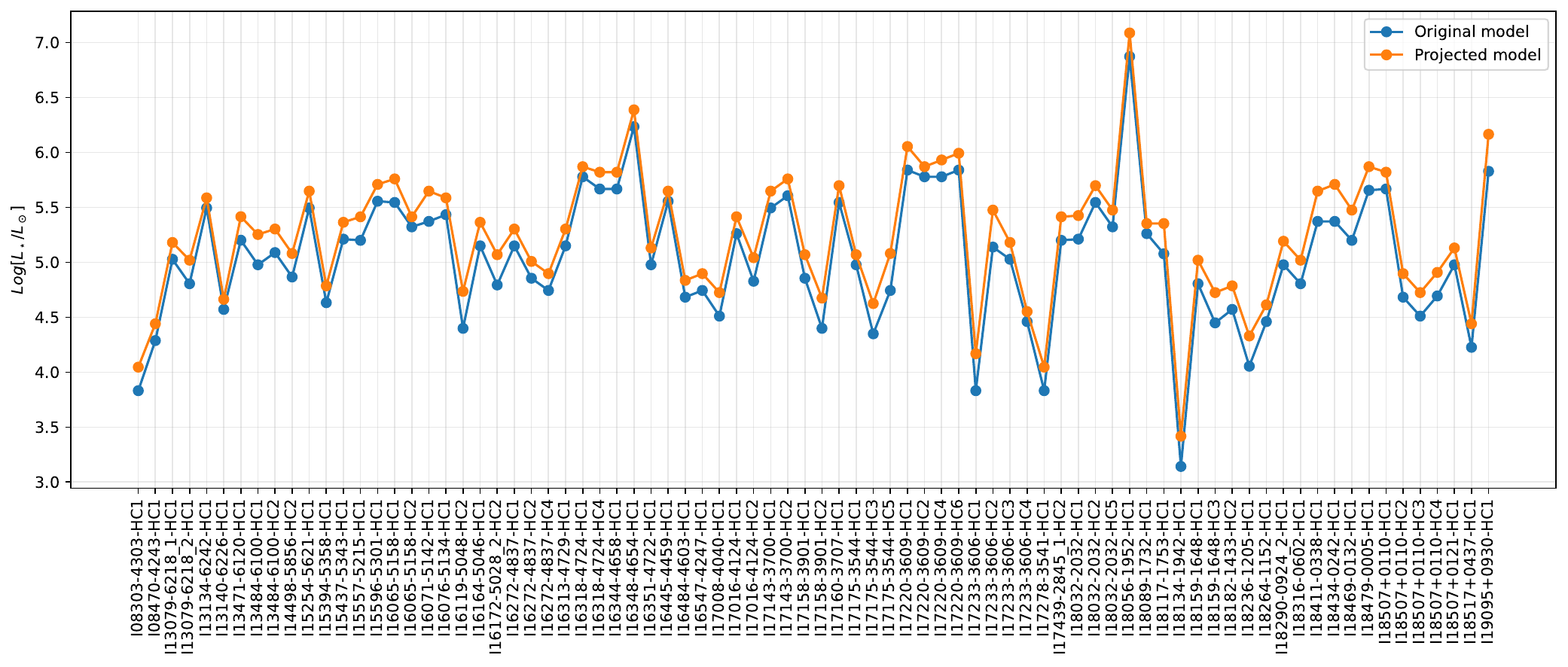}
    \caption{Comparison of protostellar luminosities derived using the original \radmc{} model (Figure~\ref{fig:workflow} solid lines) and the projection-corrected model (Figure~\ref{fig:workflow} dashed lines) for all cores. The orange points/lines indicate the projection-corrected model, while the blue points/lines indicate the original model.
    }
    \label{fig:projection_model1}
\end{figure}

\added{
In our main analysis, the model temperature profile $T_{\rm mod}$ was originally interpreted in spherical radius $r$, while the observationally derived temperatures are measured in projected cylindrical annuli. To reduce the bias introduced by this inconsistency, we implemented a projection-corrected model temperature profile (dashed lines in Figure~\ref{fig:workflow}). For each core, we integrate the three-dimensional \radmc{} temperature field along the line of sight to obtain a mass-weighted mean temperature at each cylindrical radius $R$ in the plane of the sky:
\begin{equation}
T_{\rm mod-proj}(R) = \frac{\int T_{\rm mod}(r)\, n(r)\, dz}{\int n(r)\, dz},
\end{equation}
where $r^2 = R^2 + z^2$, $n(r)$ is the \ch3cn{} volume density (derived with the same method described in Section \ref{sec: Density profile}, using H$_2$ volume density as an alternative for the four cores with failed \ch3cn{} Plummer-like function fits), and the integral is performed along the line of sight. 
The resulting $T_{\rm mod-proj}$ and observational $T_{\rm obs}$ are then used to minimize $\chi^2$ for constraining the luminosity. The luminosity derived with the projection-corrected \radmc{} model is also listed in the machine-readable Table~\ref{tab: hot cores}.}

\added{
Figure~\ref{fig:projection_model1} shows the comparison between protostellar luminosities derived using the original \radmc{} model and the projection-corrected \radmc{} model. The differences are small, with a mean $\Delta (Log [L_\star]) = 0.20$, a minimum $\Delta (Log [L_\star]) = 0.09$, and a maximum $\Delta (Log [L_\star]) = 0.34$. This indicates that the projection correction has a minor effect on the derived luminosities and does not impact any of the conclusions presented in the main text.
}

\section{Possible Distance-related Bias}
\label{Possible Distance-related Bias}
\restartappendixnumbering

\begin{figure}[ht!]
    \centering
    \includegraphics[width=18cm]{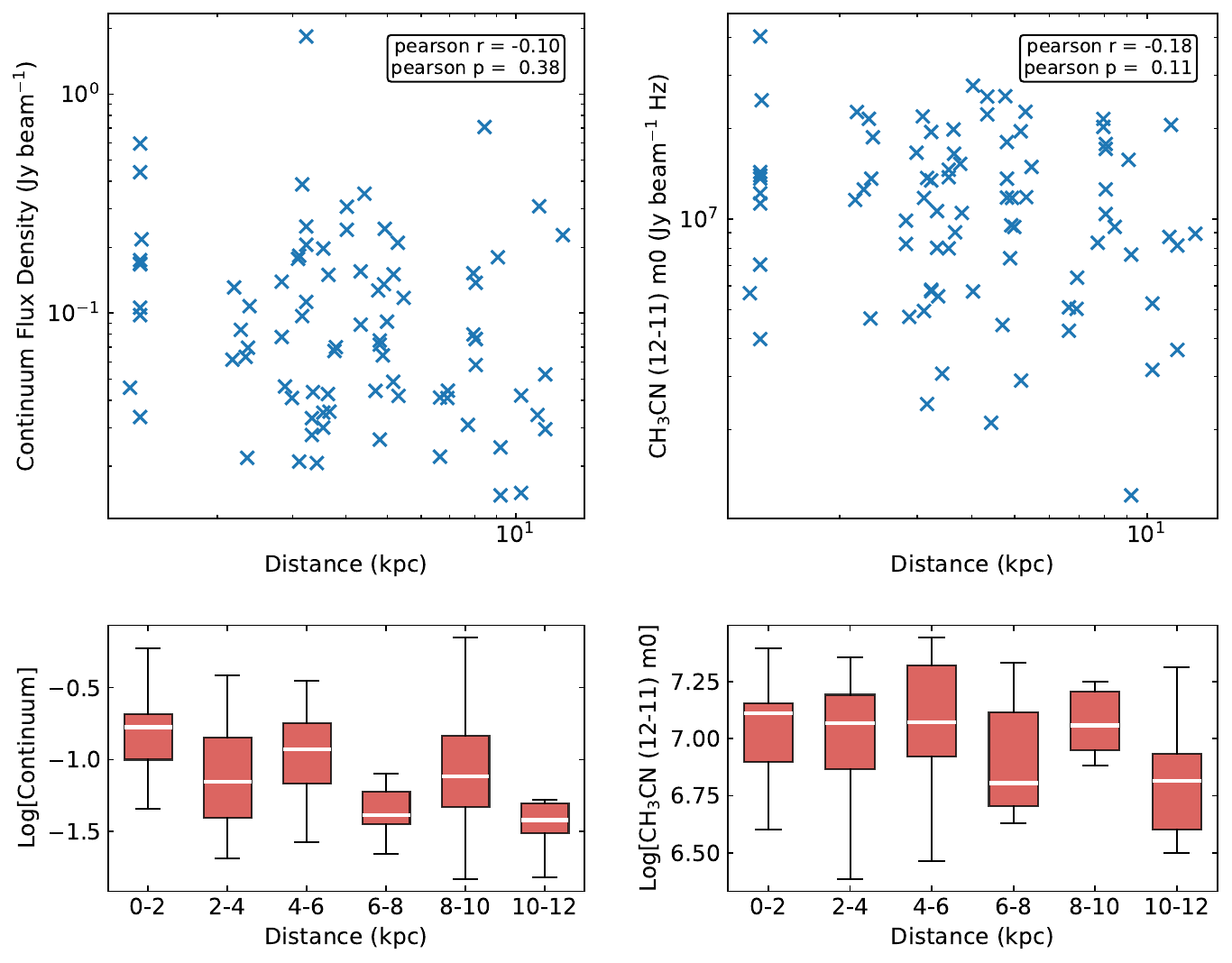}
    \caption{Dependence of the continuum flux density and \ch3cn\,(12--11, K=0--4) line integrated intensity on the distance. The top panels show the 1.3~mm continuum (peak) flux density versus the distance (left) and the \ch3cn\,(12--11,K=0--4) line integrated intensity versus the distance (right) for all cores. The corresponding Pearson correlation coefficients and p-values are indicated in the top-right corner of each panel.
    The bottom panels present box plots of the 1.3~mm continuum intensity (left) and  the \ch3cn\,(12--11, K=0--4) line integrated intensity (right) for cores grouped into 2 kpc distance bins.
    The boxes represent the inter-quartile range (25–75\%), while the white horizontal lines indicate the median values.
    }
    \label{fig:distance}
\end{figure}

To evaluate whether the observed properties of our sample are affected by the large range of distances, we examine the dependence of the continuum intensity and the \ch3cn\,(12--11, K=0--4) line integrated intensity (at the position of the 1.3~mm continuum peak) on source distance (see Figure~\ref{fig:distance}).
As shown in the lower panels of Figure~\ref{fig:distance}, the median continuum intensity and the median line integrated intensity remain approximately constant within 10 kpc.
The Pearson correlation analysis yields $r = -0.10$ (p = 0.38) for the continuum versus distance 
and $r = -0.18$ (p = 0.11) for the line intensity versus distance.
No systematic trend with increasing distance is observed.

\begin{figure}[ht!]
    \centering
    \includegraphics[width=18cm]{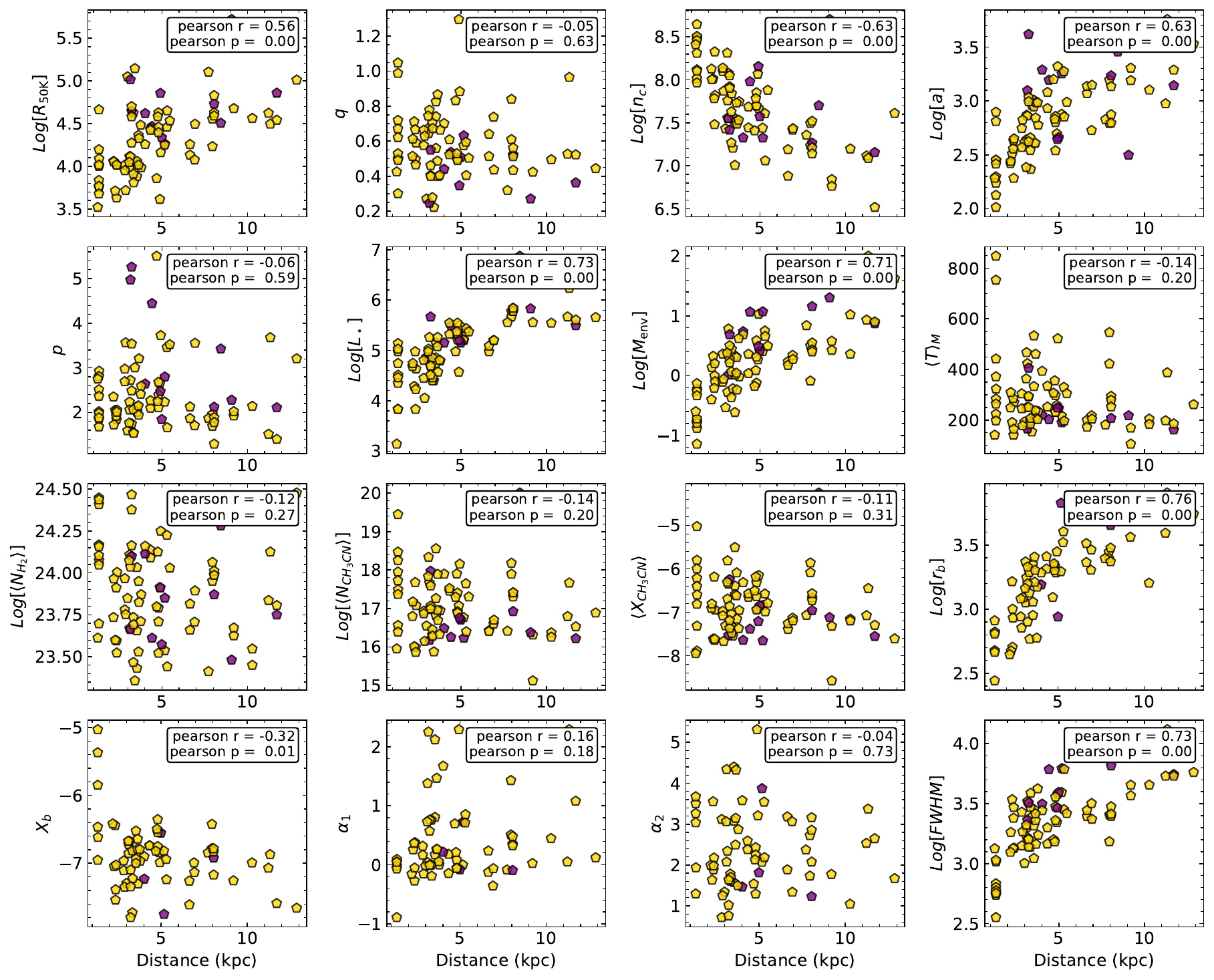}
    \caption{The scatter plots between the physical quantities shown in Figure~\ref{fig:hist} and distance. The pearson correlation coefficients and corresponding p values are labeled in the top-left corner of each panel.
    An additional physical quantity, the core size (FWHM$ = \sqrt{\theta_\mathrm{mag}\theta_\mathrm{min}}$, where $\theta_\mathrm{mag}$ and $\theta_\mathrm{min}$ are listed in Table~\ref{tab: hot cores}), is displayed in the last panel.
    }
    \label{fig:distance1}
\end{figure}

\begin{figure}[ht!]
    \centering
    \includegraphics[width=18cm]{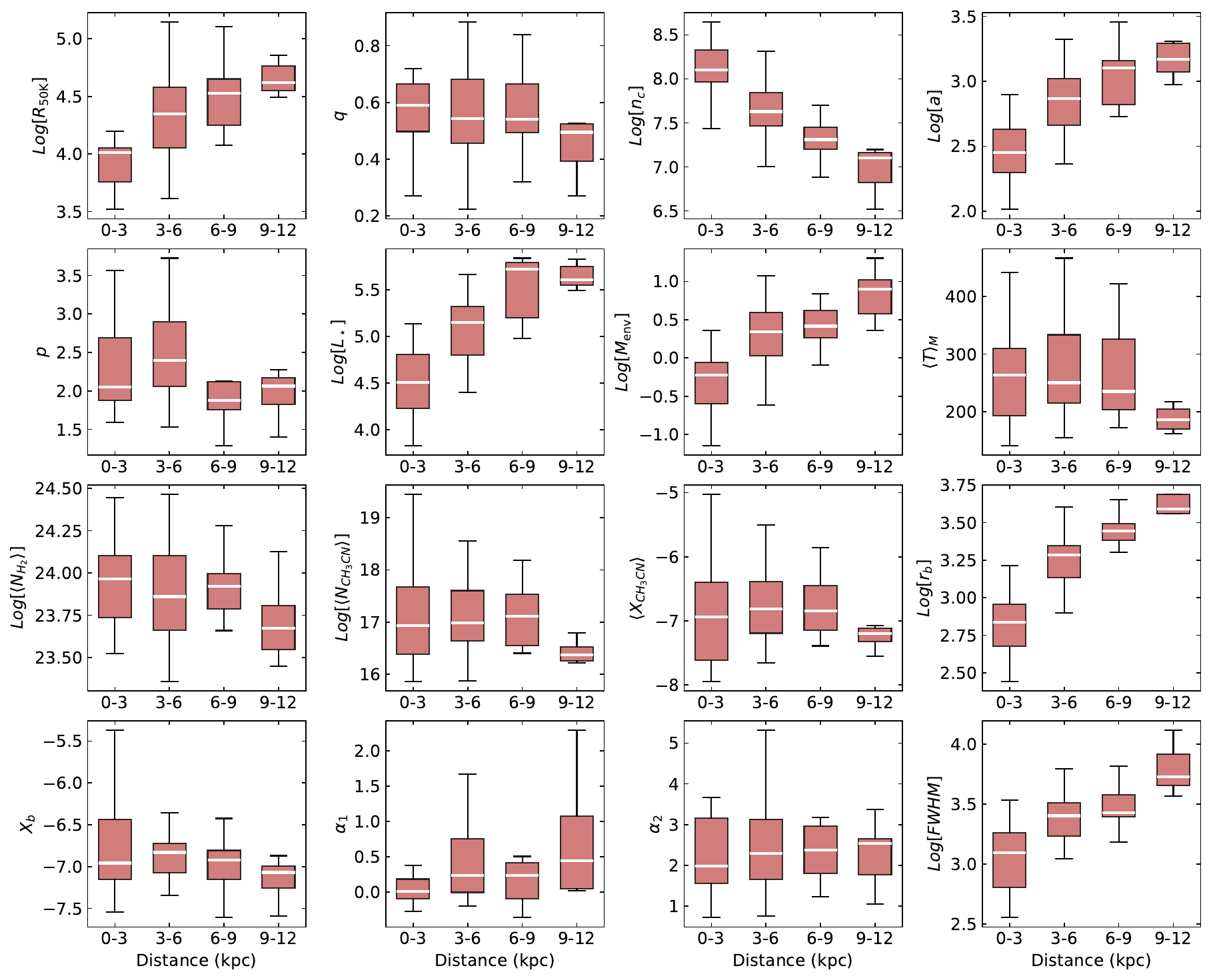}
    \caption{The corresponding box plots of Figure~\ref{fig:distance1}. The boxes represent the inter-quartile range (25–75\%), while the white horizontal lines indicate the median values.
    }
    \label{fig:distance2}
\end{figure}

In addition, we also examined the scatter plots between the physical quantities shown in Figure~\ref{fig:hist} and distance. 
Several stronger correlations with distance are found in Figure~\ref{fig:distance1} and \ref{fig:distance2}, such as luminosity $Log[L_\star]$, core size $Log[\mathrm{FWHM}]$, and envelope mass $Log[M_\mathrm{env}]$.

\section{A simple chemical model of hot core}
\label{sec:A simple chemical model of hot core}
\restartappendixnumbering

\begin{figure*}[ht!]
\includegraphics[width=\textwidth]{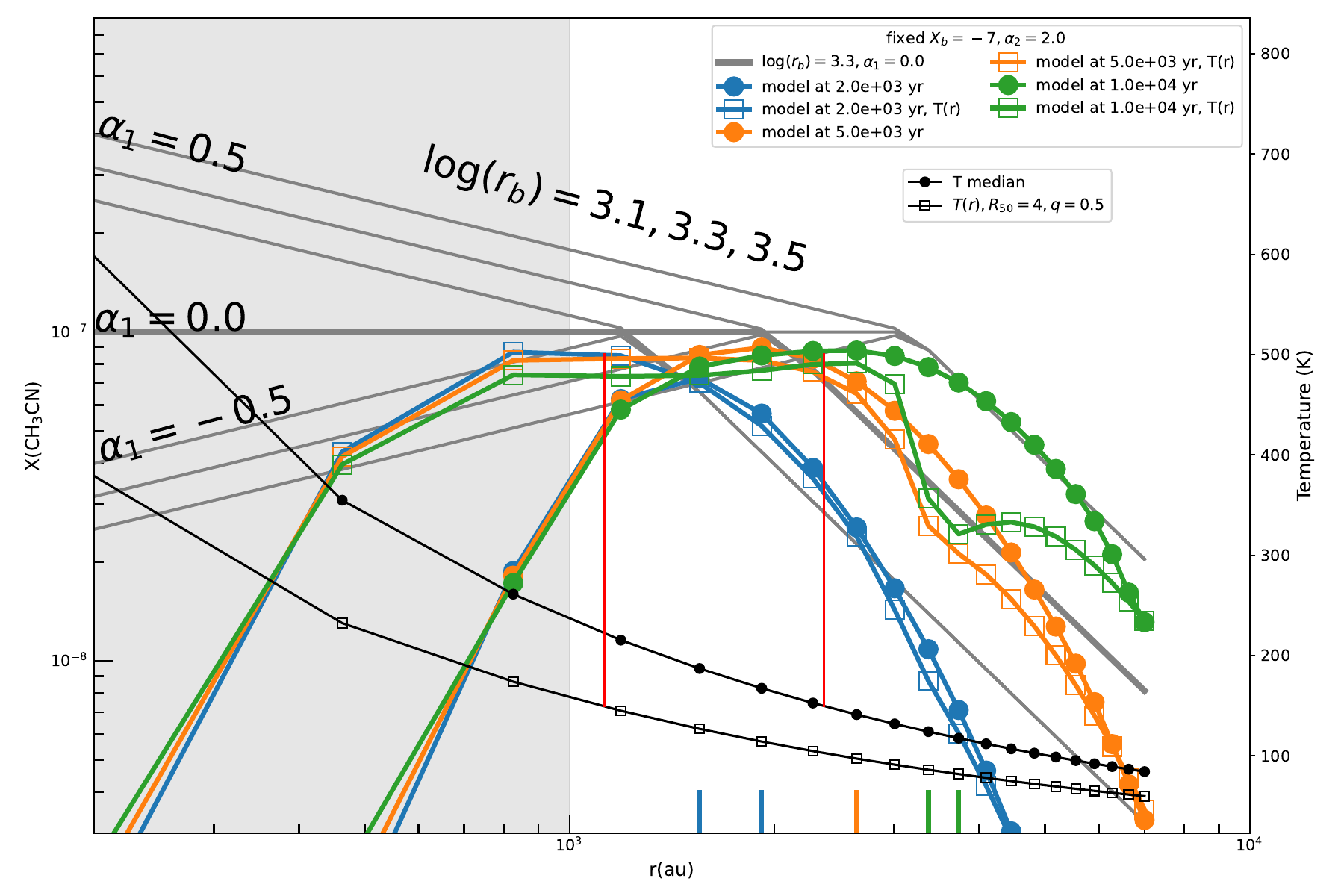}
\caption{Model results (color lines with filled circles or squares) with fitted abundance profiles(solid gray lines with different parameters). The colorful lines with points and squares are for models { Tmeadian} and { Tfunc} respectively. 
The black lines show temperature profiles used in our two models: { Tmedian} and { Tfunc} (right axis). Temperature of 150\,K are marked for the two temperature profiles with red vertical lines. At bottom, small vertical lines indicate radii with modeled abundance of $0.85X_{\rm peak}$, serving as rough estimation of break radii from models. NOTE: $T(r)$=Tfunc}
\label{fig:hc}
\end{figure*}

To study the \ce{CH3CN} abundance, the \texttt{GGCHEMPY} code \citep{Ge2022} with a gas-grain reaction network (\texttt{network2.txt}) is used to model hot core chemistry. 
For the physical structure of a hot core, the density, temperature and extinction ($A_V = N_\mathrm{H_2}(r)/2.21\times10^{21}$) are adopted as the median values from observations.
To verify the effect of temperature, another temperature profile is used with $R_\mathrm{50K}=10{,}000$ au and $q=0.5$ (Equation~\ref{eq: T_pow-law}), which are typical values according to observations and produces lower temperatures toward the center (down to 500\,K).
Finally, two hot cores models are used with only different temperature profiles: { Tmedian} and { Tfunc}. 
Other parameters are fixed as typical ones (e.g. dust-to-gas ration of 0.01, $\chi=1.0\chi_0$, $\zeta_{\rm CR}=1.3\times 10^{-17}$\,s$^{-1}$). The initial abundances are same to \citet{Ge2020}.

The model results are shown in Figure~\ref{fig:hc}. To constrain the models, fitted abundance profiles from observations are represented by solid gray lines with varying parameters labeled in the legend, holding fixed $X_b = -7$ and $\alpha_2 = 2.0$, while $\alpha_1$ takes values of $-0.5$, $0.0$, and $0.5$, and $\log(r_b)$ ranges of $3.1$, 3.3 and $3.5$.  

Modeled \ce{CH3CN} abundances at evolutionary stages of $2\times 10^3$, $5 \times 10^3$, and $10^4$ years reproduce the fitted abundances at larger radii ($r > 1{,}000$ AU). These results support $\alpha_2 = 2.0$, but they show limited sensitivity to temperature distribution at large radii ($r > 1{,}000$ AU). Furthermore, the break radius defined by the observed (fitted) abundance profiles (gray lines) aligns well with the decreasing points of the modeled abundances (lines with points) at temperatures $\sim 120$–$200$ K (e.g., the red line in the figure marks $150$ K at the model time of $5 \times 10^3$ years). A rough estimation of the break radius at $r$ with $r_b = r[0.85X_{\rm peak}]$ from the two models yields $r_b \sim 1{,}500$–$3{,}700$ AU, consistent with the observed values. Thus, the break radius likely represents the boundary between an outer warm region ($< 120$ K) and an inner hot region ($> 200$ K), which is sensitive to the evolutionary stage (e.g., model time) of the hot core.  

Significant discrepancies exist between the modeled and fitted abundances at small radii ($r < 1{,}000$ AU). This can be partly explained by temperature profile effects, supported by the {Tfunc} model (lines with empty squares). Compared to the { Tmedian} model (lines with points), { Tfunc} yields higher \ce{CH3CN} abundances at lower temperatures (e.g., $400$ K at $r = 500$ AU, $X = 1 \times 10^{-8}$), whereas { Tmedian} produces abundances as low as $5 \times 10^{-9}$. If beam effects (gray area in Fig.~\ref{fig:hc}) are considered, the model can reproduce the fitted abundances for $\alpha_1 = -0.5$ and $\alpha_1 = 0.0$. For higher observed abundances toward the center with $\alpha_1 > 0.5$ and elevated temperatures, an improved model incorporating the destruction of refractory organics as discussed by \citet{Nazari+2023} is needed to enhance the \ce{CH3CN} abundance. However, our current model with lower temperatures (e.g., { Tfunc}) slows the destruction reactions of \ce{CH3CN}, without including the enhanced reaction pathways described in \citet{Nazari+2023}.  

In conclusion, the inner abundance of \ce{CH3CN} is sensitive to the temperature profiles, and the break radius ($r_b$) of the abundance profile likely marks the boundary between the inner hot region ($> 200$ K) and the outer warm region ($< 120$ K). Within the hot core’s inner region, \ce{CH3CN} is primarily formed via efficient gas-phase ion-molecule reactions following the destruction of refractory organics (e.g., \ce{CH3+ + HCN -> CH3CNH+ -> CH3CN}, \citealt{Nazari+2023}). In the outer warm region, its abundance is dominated by desorption of ice-mantle precursors and influenced by FUV-driven photochemistry. With lower density and higher FUV photons, \ce{CH3CN} abundance decreases toward the outer region. If the destruction of refractory organics under high temperatures is accounted for, the break radius may serve as a constraint on the boundary between warm and hot regions.

The chemical model used to reproduce the \ch3cn{} abundance profile is expected to incorporate more physical processes in the future, such as the prestellar collapse phase, protostellar heating, and the internal radiation field from the central protostar. For example, \citet{2026arXiv260100731T} developed an improved chemical model specifically tailored to protostellar environments.

\section{The correlation results of some parameter pairs}
\restartappendixnumbering

This section summarizes the correlation analysis for parameter pairs with absolute Pearson correlation coefficients exceeding 0.45, including the best-fit linear slope and intercept, with their uncertainties, along with the Pearson and Spearman correlation coefficients ($r_p$ and $r_s$) and their associated p-value ($p_p$ and $p_s$).
Table~\ref{tab:correlation} is sorted in descending order of the Pearson correlation coefficient ($r_p$). Note that $r_p^\mathrm{lim}$ and $p_p^\mathrm{lim}$ are the Pearson coefficients and the corresponding p-values for the distance-limited sample.

\startlongtable
\tabletypesize{\scriptsize}
\begin{deluxetable*}{llrrrrrrrrrr}
\tablecaption{The correlation results of some parameter pairs. \label{tab:correlation}}
\tablehead{
\colhead{x} & \colhead{y} & \colhead{slope} & \colhead{slope error} & \colhead{intercept} & \colhead{intercept error} & \colhead{$r_p$} & \colhead{$Log[p_p]$} & \colhead{$r_s$} & \colhead{$Log[p_s]$}& \colhead{$r_p^\mathrm{lim}$} & \colhead{$Log[p_p^\mathrm{lim}]$}
}
\colnumbers
\startdata
$Log[\langle T \rangle_M]$ & $Log[\langle N_\mathrm{CH_3CN} \rangle]$ & 4.96 & 0.17 & 5.06 & 0.41 & 0.96 & -43.94 & 0.95 & -42.41 & 0.95 & -27.87 \\
$Log[\langle N_\mathrm{CH_3CN} \rangle]$ & $\langle X_\mathrm{CH_3CN} \rangle$ & 0.80 & 0.03 & -20.55 & 0.50 & 0.95 & -41.46 & 0.92 & -34.41 & 0.94 & -26.27 \\
$Log[\langle T \rangle_M]$ & $\langle X_\mathrm{CH_3CN} \rangle$ & 4.25 & 0.23 & -17.13 & 0.56 & 0.90 & -29.78 & 0.87 & -25.56 & 0.88 & -18.60 \\
$\langle X_\mathrm{CH_3CN} \rangle$ & $X_b$ & 0.81 & 0.07 & -1.43 & 0.50 & 0.76 & -13.31 & 0.78 & -13.99 & 0.74 & -8.44 \\
$Log[L_\star]$ & $Log[M_\mathrm{Jeans}]$ & 0.32 & 0.03 & -0.63 & 0.15 & 0.75 & -15.07 & 0.67 & -11.05 & 0.69 & -8.40 \\
$Log[L_\star]$ & $Log[r_b]$ & 0.56 & 0.06 & 0.37 & 0.29 & 0.74 & -11.74 & 0.78 & -13.81 & 0.55 & -4.07 \\
$Log[L_\star]$ & $Log[M]$ & 1.01 & 0.10 & -4.80 & 0.49 & 0.70 & -12.29 & 0.70 & -12.29 & 0.49 & -3.87 \\
$Log[\langle N_\mathrm{H_2}\rangle]$ & $Log[\langle N_\mathrm{CH_3CN} \rangle]$ & 4.06 & 0.45 & -79.83 & 10.74 & 0.69 & -12.18 & 0.74 & -14.68 & 0.70 & -8.64 \\
$Log[\langle N_\mathrm{CH_3CN} \rangle]$ & $X_b$ & 0.56 & 0.06 & -16.49 & 1.10 & 0.68 & -9.47 & 0.62 & -7.65 & 0.67 & -6.42 \\
$Log[\langle T \rangle_M]$ & $Log[\langle N_\mathrm{H_2}\rangle]$ & 1.97 & 0.20 & 19.12 & 0.50 & 0.67 & -11.55 & 0.72 & -13.77 & 0.68 & -7.96 \\
$Log[a]$ & $Log[r_b]$ & 1.07 & 0.12 & 0.17 & 0.34 & 0.67 & -9.16 & 0.65 & -8.56 & 0.45 & -2.76 \\
$Log[M]$ & $Log[r_b]$ & 0.48 & 0.06 & 3.08 & 0.03 & 0.67 & -9.23 & 0.61 & -7.32 & 0.45 & -2.80 \\
$Log[a]$ & $Log[M]$ & 2.12 & 0.23 & -5.76 & 0.66 & 0.67 & -11.01 & 0.61 & -8.97 & 0.54 & -4.64 \\
$q$ & $Log[\langle T \rangle_M]$ & 0.76 & 0.08 & 1.99 & 0.05 & 0.66 & -10.66 & 0.58 & -7.98 & 0.62 & -6.40 \\
$Log[\langle T \rangle_M]$ & $X_b$ & 4.37 & 0.59 & -17.52 & 1.45 & 0.65 & -8.63 & 0.58 & -6.63 & 0.66 & -6.18 \\
$Log[L_\star]$ & $Log[a]$ & 0.46 & 0.05 & 0.52 & 0.26 & 0.65 & -10.22 & 0.59 & -8.02 & 0.56 & -5.15 \\
$q$ & $\langle X_\mathrm{CH_3CN} \rangle$ & 4.71 & 0.60 & -9.56 & 0.36 & 0.64 & -9.89 & 0.59 & -8.28 & 0.60 & -5.89 \\
$q$ & $Log[\langle N_\mathrm{CH_3CN} \rangle]$ & 6.02 & 0.80 & 13.59 & 0.48 & 0.63 & -9.53 & 0.55 & -7.07 & 0.57 & -5.30 \\
$Log[R_\mathrm{50K}]$ & $Log[r_b]$ & 0.83 & 0.10 & -0.34 & 0.44 & 0.61 & -7.38 & 0.66 & -9.06 & 0.48 & -3.10 \\
$q$ & $\alpha_2$ & 8.88 & 1.52 & -2.95 & 0.93 & 0.58 & -6.48 & 0.51 & -4.94 & 0.60 & -5.05 \\
$Log[L_\star]$ & $Log[R_\mathrm{50K}]$ & 0.54 & 0.07 & 1.60 & 0.36 & 0.55 & -6.93 & 0.59 & -8.01 & 0.40 & -2.66 \\
$q$ & $Log[M_\mathrm{Jeans}]$ & 1.37 & 0.18 & 0.20 & 0.11 & 0.50 & -5.60 & 0.44 & -4.28 & 0.54 & -4.73 \\
$q$ & $X_b$ & 4.69 & 0.92 & -9.73 & 0.56 & 0.49 & -4.65 & 0.43 & -3.61 & 0.52 & -3.74 \\
$Log[R_\mathrm{50K}]$ & $Log[M_\mathrm{env}]$ & 2.03 & 0.30 & -8.49 & 1.30 & 0.48 & -5.30 & 0.51 & -5.82 & 0.27 & -1.34 \\
$Log[a]$ & $p$ & 4.40 & 0.80 & -10.06 & 2.28 & 0.48 & -5.17 & 0.41 & -3.89 & 0.66 & -7.28 \\
$Log[L_\star]$ & $\langle X_\mathrm{CH_3CN} \rangle$ & 1.20 & 0.16 & -12.86 & 0.83 & 0.47 & -5.03 & 0.36 & -2.96 & 0.60 & -5.81 \\
$Log[L_\star]$ & $Log[n_c]$ & -0.55 & 0.08 & 10.47 & 0.41 & -0.48 & -5.13 & -0.50 & -5.53 & -0.38 & -2.42 \\
$Log[n_c]$ & $Log[r_b]$ & -0.68 & 0.09 & 8.37 & 0.68 & -0.58 & -6.59 & -0.60 & -7.12 & -0.30 & -1.34 \\
$Log[R_\mathrm{50K}]$ & $q$ & -0.34 & 0.04 & 2.04 & 0.19 & -0.59 & -8.25 & -0.66 & -10.52 & -0.67 & -7.81 \\
$Log[n_c]$ & $Log[a]$ & -0.69 & 0.07 & 8.15 & 0.53 & -0.68 & -11.67 & -0.67 & -11.04 & -0.65 & -7.15 \\
\enddata
\tablecomments{$Log[0.05]\simeq-1.30$}
\end{deluxetable*}

\section{The parameters for the identified hot cores.}
\restartappendixnumbering

This section presents Table \ref{tab: hot cores}, which lists some parameters of the hot cores identified in Section \ref{sec: The identification of internally heated hot core}. These parameters include hot core target (Name, column 1), right ascension (R.A., column 2), declination (Decl., column 3), FWHM$_{\mathrm{maj}}$ ($\theta_\mathrm{maj}$, column 4) along the major axis of core, FWHM$_{\mathrm{min}}$ ($\theta_\mathrm{min}$, column 5) along the minor axis of core, the position angle of core (PA, column 6), radius where the temperature decreases to 50\,K ($R_{\mathrm{50K}}$, column 7), the temperature pow-law index ($q$, column 8), H$_2$ number density at the core center ($n_c$, column 9), the density flat radius ($a$, column 10), the density index ($p$, column 11), the embedded protostellar luminosity ($L_\star$, column 12), the protostellar envelope mass ($M_\mathrm{env}$, column 13),  the mass-averaged temperature ($\langle T \rangle_{M}$, column 14), the core-averaged H$_2$ column density ($\langle N_{\mathrm{H}_2} \rangle$, column 15), the core-averaged \ch3cn{} column density ($\langle N_{\mathrm{CH_3CN}} \rangle$, column 16), and the core-averaged \ch3cn{} abundance ($\langle X_{\mathrm{CH_3CN}} \rangle = \langle\mathrm{Log}[N_{\mathrm{CH_3CN}}/N_{\mathrm{H}_2}]\rangle$, column 17). \added{The parameters $r_b$, $X_b$, $\alpha_1$, $\alpha_2$, Jeans mass, projection-corrected luminosity, distance, the mass and the bolometric luminosity of clump can be found in the machine-readable table.}

\begin{longrotatetable}
\digitalasset
\movetabledown=20mm
\begin{deluxetable*}{lrrrrrccccccrrccc}
\colnumbers
\tabletypesize{\scriptsize}
\tablecaption{The parameters for 83 identified hot cores. Values in parentheses indicate the corresponding uncertainties. \label{tab: hot cores}}
\tablehead{
\colhead{Name \tablenotemark{a}} & \colhead{R.A.\tablenotemark{c}} & \colhead{Decl.\tablenotemark{c}} &
\colhead{$\theta_\mathrm{maj}$} & \colhead{$\theta_\mathrm{min}$} & \colhead{PA} &
\colhead{$R_{\mathrm{50K}}$} & \colhead{$q$} &
\colhead{Log[$n_c$]} & \colhead{$a$} & \colhead{$p$} &
\colhead{Log[$L_\star$]} &
\colhead{$M_\mathrm{env}$} & \colhead{$\langle T \rangle_{M}$ \tablenotemark{b}} &
\colhead{Log[$\langle N_{\mathrm{H}_2} \rangle$]} & \colhead{Log[$\langle N_{\mathrm{CH_3CN}} \rangle$]} & \colhead{$\langle X_{\mathrm{CH_3CN}} \rangle$}
\\
\colhead{} & \colhead{(deg)} & \colhead{(deg)} &
\colhead{(au)} & \colhead{(au)} & \colhead{(deg)} &
\colhead{(au)} & \colhead{} &
\colhead{(cm$^{-3}$)} & \colhead{(au)} & \colhead{} &
\colhead{($L_\odot$)} &
\colhead{($M_\odot$)} & \colhead{(K)} &
\colhead{(cm$^{-2}$)} & \colhead{(cm$^{-2}$)} & \colhead{}
}
\startdata
I08303-4303-HC1 & 128.03601 & -43.22934 & 3298 & 1744 & 120 & 4258(121) & 0.50(0.02) & 7.48(0.05) & 427(52) & 1.84(0.05) & 3.83 & 0.99 & 141 & 23.60(0.15) & 15.99(0.44) & -7.64(0.43) \\
I08470-4243-HC1 & 132.19914 & -42.90778 & 4247 & 2757 & 58 & 5141(310) & 0.71(0.03) & 8.01(0.13) & 323(85) & 2.05(0.12) & 4.29 & 2.12 & 191 & 23.60(0.22) & 16.01(0.71) & -7.61(0.49) \\
I13079-6218\_1-HC1 & 197.80724 & -62.57809 & 2558 & 1559 & 121 & 12962(441) & 0.67(0.01) & 7.76(0.08) & 738(151) & 2.23(0.16) & 5.03 & 2.27 & 467 & 24.11(0.21) & 18.33(0.78) & -5.81(0.72) \\
I13079-6218\_2-HC1 & 197.79370 & -62.57737 & 1942 & 1436 & 127 & 9614(2227) & 0.64(0.08) & 7.56(0.05) & 438(69) & 2.06(0.10) & 4.80 & 0.53 & 249 & 23.67(0.14) & 16.97(0.16) & -6.69(0.17) \\
I13134-6242-HC1 & 199.18003 & -62.97575 & 4159 & 3492 & -160 & 14484(954) & 0.88(0.05) & 7.67(0.04) & 2094(409) & 3.73(0.62) & 5.49 & 10.68 & 522 & 24.25(0.18) & 18.15(0.94) & -6.15(0.90) \\
I13140-6226-HC1 & 199.31467 & -62.70648 & 2308 & 2116 & -163 & 4098(443) & 1.29(0.24) & 8.07(0.11) & 497(91) & 2.65(0.13) & 4.57 & 1.85 & 258 & 23.91(0.20) & 17.07(0.32) & -6.92(0.21) \\
I13471-6120-HC1* & 207.67420 & -61.58625 & 6738 & 5736 & 68 & 18287(1403) & 0.63(0.04) & 7.32(0.08) & 1794(296) & 2.80(0.17) & 5.20 & 11.76 & 190 & 23.85(0.16) & 16.23(0.52) & -7.65(0.51) \\
I13484-6100-HC1 & 207.99320 & -61.26143 & 2731 & 2249 & 151 & 13550(962) & 0.64(0.04) & 7.19(0.07) & 1174(279) & 2.13(0.21) & 4.98 & 1.72 & 200 & 23.82(0.07) & 16.42(0.16) & -7.39(0.17) \\
I13484-6100-HC2 & 207.99269 & -61.26136 & 3174 & 2436 & -145 & 18031(1611) & 0.51(0.04) & 6.88(0.04) & 1372(228) & 1.86(0.11) & 5.09 & 1.49 & 172 & 23.66(0.08) & 16.40(0.12) & -7.26(0.19) \\
I14498-5856-HC2 & 223.42764 & -59.14807 & 1066 & 951 & 73 & 112123(44704) & 0.27(0.03) & 7.43(0.05) & 412(62) & 1.59(0.05) & 4.87 & 0.29 & 193 & 23.85(0.05) & 16.55(0.05) & -7.31(0.07) \\
I15254-5621-HC1* & 232.33066 & -56.52281 & 8269 & 4501 & 87 & 29267(4308) & 0.54(0.04) & 7.98(0.06) & 1567(257) & 4.44(0.51) & 5.49 & 11.64 & 202 & 23.61(0.51) & 16.25(0.27) & -7.38(0.52) \\
I15394-5358-HC1 & 235.81935 & -54.12075 & 1671 & 1086 & 113 & 10370(743) & 0.59(0.02) & 7.82(0.10) & 446(135) & 2.05(0.22) & 4.63 & 0.79 & 270 & 24.00(0.08) & 17.08(0.24) & -6.94(0.22) \\
I15437-5343-HC1 & 236.88635 & -53.87741 & 5655 & 2734 & 176 & 17806(2189) & 0.57(0.04) & 7.44(0.18) & 755(307) & 2.24(0.28) & 5.21 & 2.53 & 217 & 23.54(0.21) & 16.77(0.28) & -6.84(0.18) \\
I15557-5215-HC1 & 239.91962 & -52.39107 & 3893 & 2596 & 108 & 11885(878) & 0.74(0.04) & 7.42(0.21) & 534(233) & 1.70(0.10) & 5.20 & 2.19 & 217 & 23.71(0.14) & 16.60(0.25) & -7.13(0.25) \\
I15596-5301-HC1 & 240.88390 & -53.15840 & 5013 & 4078 & 159 & 47504(14945) & 0.42(0.06) & 6.84(0.09) & 1553(378) & 1.93(0.14) & 5.56 & 3.78 & 169 & 23.63(0.09) & 16.31(0.14) & -7.32(0.16) \\
I15596-5301-HC2 & 240.88343 & -53.15850 & 4039 & 3361 & 175 & nan(nan) & nan(nan) & 6.76(0.03) & 2025(333) & 2.02(0.16) & nan & 2.66 & 106 & 23.67(0.06) & 15.12(0.21) & -8.57(0.22) \\
I16065-5158-HC1 & 242.58295 & -52.10213 & 2639 & 1507 & 137 & 27577(2333) & 0.61(0.04) & 7.69(0.03) & 854(64) & 2.10(0.05) & 5.54 & 2.29 & 335 & 24.14(0.11) & 17.46(0.26) & -6.65(0.37) \\
I16065-5158-HC2 & 242.58326 & -52.10192 & 1689 & 1263 & 99 & 26434(4543) & 0.52(0.04) & 7.44(0.05) & 1204(143) & 2.27(0.08) & 5.32 & 1.09 & 252 & 24.09(0.09) & 17.52(0.15) & -6.57(0.23) \\
I16071-5142-HC1 & 242.74892 & -51.83970 & 3290 & 2427 & 60 & 28721(3194) & 0.59(0.03) & 7.61(0.10) & 1884(701) & 3.46(0.92) & 5.37 & 5.60 & 330 & 24.22(0.11) & 17.65(0.37) & -6.57(0.42) \\
I16076-5134-HC1 & 242.86058 & -51.69928 & 7635 & 4907 & 110 & 44792(7006) & 0.40(0.03) & 7.06(0.23) & 713(377) & 1.66(0.14) & 5.43 & 4.55 & 197 & 23.44(0.13) & 16.40(0.43) & -6.99(0.40) \\
I16119-5048-HC2 & 243.93908 & -50.93157 & 3388 & 1781 & -139 & 139870(26513) & 0.22(0.01) & 7.21(0.05) & 912(228) & 3.01(0.55) & 4.40 & 0.61 & 154 & 23.36(0.20) & 15.87(0.29) & -7.54(0.28) \\
I16164-5046-HC1* & 245.04608 & -50.88727 & 3258 & 3029 & 176 & 41553(6176) & 0.44(0.03) & 7.33(0.09) & 1947(475) & 2.64(0.30) & 5.15 & 5.45 & 219 & 24.11(0.11) & 16.48(0.42) & -7.64(0.45) \\
I16164-5046-HC2 & 245.04607 & -50.88766 & 2807 & 2741 & 123 & 18025(1201) & 0.70(0.03) & nan(nan) & nan(nan) & nan(nan) & nan & 4.87 & 393 & 24.16(0.15) & 17.90(0.49) & -6.31(0.59) \\
I16172-5028\_2-HC2 & 245.24889 & -50.58498 & 1790 & 1675 & 72 & 9647(3076) & 0.66(0.14) & 7.75(0.16) & 598(211) & 2.59(0.33) & 4.79 & 0.96 & 235 & 23.87(0.11) & 16.95(0.17) & -6.95(0.20) \\
I16272-4837-HC1 & 247.74485 & -48.73167 & 2144 & 1979 & 59 & 10459(504) & 0.82(0.03) & 7.50(0.16) & 884(404) & 2.13(0.33) & 5.15 & 2.15 & 533 & 24.03(0.17) & 18.56(0.65) & -5.51(0.74) \\
I16272-4837-HC2 & 247.73869 & -48.72781 & 3531 & 2276 & 173 & 7589(2129) & 0.74(0.14) & 7.55(0.25) & 408(229) & 2.01(0.27) & 4.85 & 1.03 & 217 & 23.43(0.18) & 16.81(0.49) & -6.59(0.39) \\
I16272-4837-HC4 & 247.74428 & -48.73090 & 1288 & 951 & 118 & 11408(798) & 0.56(0.02) & 7.01(0.06) & 1019(391) & 2.01(0.42) & 4.74 & 0.24 & 333 & 23.66(0.11) & 17.26(0.37) & -6.39(0.46) \\
I16313-4729-HC1* & 248.72675 & -47.59372 & 5180 & 2987 & 145 & 21684(13128) & 0.51(0.14) & 7.57(0.24) & 458(238) & 1.84(0.18) & 5.15 & 2.57 & 246 & 23.57(0.24) & 16.71(0.45) & -6.85(0.42) \\
I16318-4724-HC1 & 248.89146 & -47.51994 & 2858 & 2272 & -175 & 35764(3361) & 0.61(0.03) & 7.23(0.22) & 1574(873) & 1.96(0.30) & 5.78 & 3.16 & 546 & 24.06(0.12) & 18.18(0.36) & -5.86(0.46) \\
I16318-4724-HC4 & 248.89123 & -47.51999 & 1804 & 1293 & 153 & 17036(753) & 0.84(0.03) & 7.49(0.53) & 620(535) & 1.69(0.09) & 5.67 & 0.81 & 423 & 23.95(0.08) & 17.90(0.19) & -6.08(0.27) \\
I16344-4658-HC1 & 249.53936 & -47.08325 & 6634 & 4355 & 80 & 41910(14818) & 0.53(0.10) & 7.12(0.30) & 945(688) & 1.51(0.15) & 5.67 & 8.54 & 198 & 23.84(0.07) & 16.79(0.10) & -7.07(0.10) \\
I16348-4654-HC1 & 249.62344 & -47.00999 & 16241 & 10592 & 49 & 31263(1541) & 0.96(0.04) & 7.09(0.04) & 5725(1041) & 3.68(0.54) & 6.23 & 99.78 & 388 & 24.12(0.17) & 17.67(0.94) & -6.45(0.94) \\
I16351-4722-HC1 & 249.71037 & -47.46683 & 1471 & 1220 & 151 & 10513(431) & 0.67(0.02) & 7.78(0.09) & 360(74) & 1.92(0.08) & 4.98 & 0.62 & 286 & 23.91(0.12) & 17.68(0.15) & -6.18(0.20) \\
I16445-4459-HC1 & 252.02143 & -45.08568 & 9863 & 7233 & 71 & 127284(50241) & 0.32(0.04) & 7.36(0.39) & 673(508) & 1.87(0.22) & 5.56 & 6.89 & 182 & 23.41(0.19) & 16.41(1.65) & -7.07(0.32) \\
I16484-4603-HC1 & 253.01942 & -46.14283 & 1682 & 922 & 68 & 11391(1456) & 0.51(0.03) & 8.13(0.12) & 263(64) & 2.35(0.17) & 4.68 & 0.40 & 249 & 23.74(0.19) & 16.80(0.22) & -6.91(0.25) \\
I16547-4247-HC1 & 254.57175 & -42.86876 & 2268 & 1468 & -143 & 11298(749) & 0.61(0.02) & 8.00(0.06) & 275(30) & 1.72(0.03) & 4.74 & 1.38 & 309 & 23.96(0.13) & 17.17(0.51) & -6.80(0.53) \\
I17008-4040-HC1 & 256.09539 & -40.73969 & 1126 & 954 & 58 & 12213(2037) & 0.50(0.04) & 8.49(0.09) & 280(59) & 2.93(0.28) & 4.51 & 0.60 & 323 & 24.07(0.21) & 17.49(0.28) & -6.60(0.22) \\
I17016-4124-HC1 & 256.29540 & -41.48524 & 4255 & 2566 & 63 & 12960(364) & 0.78(0.02) & 8.08(0.05) & 728(100) & 2.72(0.18) & 5.26 & 6.04 & 424 & 24.04(0.24) & 17.67(0.85) & -6.35(0.72) \\
I17016-4124-HC2* & 256.29668 & -41.48529 & 2560 & 2084 & 153 & 103462(115989) & 0.25(0.07) & 7.55(0.06) & 1250(444) & 4.97(1.84) & 4.83 & 1.03 & 165 & 23.66(0.37) & 16.17(0.50) & -7.53(0.77) \\
I17143-3700-HC1* & 259.43940 & -37.05362 & 6508 & 4725 & 138 & 72057(42271) & 0.36(0.08) & 7.16(0.28) & 1391(869) & 2.11(0.34) & 5.49 & 7.45 & 162 & 23.75(0.09) & 16.22(0.11) & -7.55(0.12) \\
I17143-3700-HC2 & 259.43933 & -37.05328 & 8237 & 3507 & 168 & 34753(2093) & 0.52(0.02) & 6.52(0.08) & 1943(944) & 1.40(0.17) & 5.61 & 7.96 & 186 & 23.81(0.07) & 16.53(0.27) & -7.29(0.31) \\
I17158-3901-HC1 & 259.83514 & -39.06374 & 3765 & 2310 & 152 & 8091(1858) & 0.70(0.11) & 7.63(0.19) & 348(122) & 2.06(0.14) & 4.85 & 1.16 & 196 & 23.47(0.14) & 16.31(0.70) & -7.04(0.70) \\
I17158-3901-HC2 & 259.83524 & -39.06453 & 1448 & 1304 & 172 & 42093(17959) & 0.28(0.04) & 7.27(0.06) & 456(87) & 1.55(0.06) & 4.40 & 0.43 & 185 & 23.74(0.05) & 16.45(0.22) & -7.32(0.23) \\
I17160-3707-HC1 & 259.86426 & -37.18547 & 11820 & 5859 & 142 & 36485(9387) & 0.49(0.06) & 7.20(0.33) & 1266(783) & 2.14(0.28) & 5.54 & 10.47 & 205 & 23.55(0.19) & 16.37(0.43) & -7.16(0.33) \\
I17160-3707-HC4 & 259.86250 & -37.18149 & 5038 & 4067 & 171 & nan(nan) & nan(nan) & nan(nan) & nan(nan) & nan(nan) & nan & 2.30 & 184 & 23.45(0.11) & 16.25(0.25) & -7.20(0.29) \\
I17175-3544-HC1 & 260.22256 & -35.78271 & 861 & 539 & 125 & 6830(347) & 1.05(0.05) & 8.12(0.02) & 647(85) & 2.51(0.24) & 4.98 & 0.55 & 848 & 24.45(0.05) & 19.45(0.41) & -5.02(0.44) \\
I17175-3544-HC3 & 260.22157 & -35.78315 & 1486 & 770 & 124 & 9991(884) & 0.53(0.02) & 8.51(0.13) & 198(49) & 2.01(0.10) & 4.35 & 0.74 & 283 & 24.17(0.15) & 17.37(0.40) & -6.81(0.44) \\
I17175-3544-HC5 & 260.22268 & -35.78286 & 382 & 332 & -139 & 5712(284) & 0.99(0.04) & 8.02(0.04) & 790(133) & 2.81(0.26) & 4.74 & 0.13 & 753 & 24.41(0.07) & 18.47(0.31) & -6.00(0.36) \\
I17220-3609-HC1* & 261.35513 & -36.21264 & 8752 & 4905 & 123 & 53773(13315) & 0.52(0.07) & 7.27(0.07) & 1717(261) & 2.12(0.09) & 5.84 & 14.37 & 207 & 23.87(0.15) & 16.92(0.21) & -6.96(0.25) \\
I17220-3609-HC2 & 261.35563 & -36.21226 & 3565 & 2508 & -169 & 42188(3883) & 0.51(0.03) & 7.21(0.05) & 1367(244) & 1.88(0.13) & 5.78 & 3.54 & 294 & 23.99(0.13) & 17.40(0.30) & -6.67(0.40) \\
I17220-3609-HC4 & 261.35539 & -36.21275 & 3050 & 2027 & 163 & 38929(14015) & 0.56(0.11) & 7.52(0.29) & 748(381) & 1.77(0.09) & 5.78 & 2.71 & 279 & 24.01(0.14) & 17.30(0.33) & -6.73(0.45) \\
I17220-3609-HC6 & 261.35528 & -36.21230 & 3541 & 1898 & 137 & 67529(8211) & 0.43(0.03) & 7.14(0.13) & 625(218) & 1.29(0.04) & 5.84 & 2.51 & 253 & 23.99(0.06) & 17.41(0.12) & -6.57(0.16) \\
I17233-3606-HC1 & 261.67617 & -36.15494 & 823 & 489 & 178 & 45965(16359) & 0.30(0.03) & 8.65(0.19) & 134(47) & 1.68(0.06) & 3.83 & 0.46 & 224 & 24.44(0.09) & 16.56(0.13) & -7.89(0.19) \\
I17233-3606-HC2 & 261.67723 & -36.15485 & 712 & 440 & -159 & 15700(1752) & 0.62(0.03) & 8.31(0.09) & 173(35) & 1.85(0.08) & 5.14 & 0.17 & 442 & 24.10(0.16) & 17.95(0.41) & -6.22(0.53) \\
I17233-3606-HC3 & 261.67701 & -36.15518 & 885 & 399 & -176 & 10351(2184) & 0.67(0.08) & 7.97(0.09) & 284(70) & 1.98(0.15) & 5.03 & 0.16 & 372 & 24.05(0.17) & 18.25(0.41) & -5.80(0.56) \\
I17233-3606-HC4 & 261.67664 & -36.15517 & 595 & 503 & 116 & 5840(959) & 0.72(0.08) & 8.10(0.05) & 260(41) & 1.88(0.10) & 4.46 & 0.19 & 264 & 24.16(0.14) & 17.72(0.27) & -6.40(0.39) \\
I17278-3541-HC1 & 262.80778 & -35.73574 & 618 & 523 & 176 & 4743(1306) & 0.49(0.07) & 8.46(0.60) & 103(82) & 2.36(0.24) & 3.83 & 0.07 & 198 & 23.70(0.12) & 16.39(0.19) & -7.31(0.23) \\
I17439-2845\_1-HC2 & 266.78799 & -28.77119 & 2486 & 2219 & 112 & 31153(2121) & 0.44(0.01) & 7.43(0.04) & 1404(134) & 3.55(0.18) & 5.20 & 1.87 & 204 & 23.89(0.09) & 16.70(0.11) & -7.20(0.11) \\
I18032-2032-HC1 & 271.56107 & -20.52544 & 5174 & 2500 & -167 & 28774(3774) & 0.49(0.03) & 7.85(0.30) & 721(392) & 2.68(0.44) & 5.21 & 3.87 & 232 & 23.79(0.22) & 16.92(0.47) & -6.94(0.34) \\
I18032-2032-HC2 & 271.56195 & -20.52767 & 4228 & 2105 & 108 & 42373(14264) & 0.50(0.06) & 7.70(0.22) & 633(273) & 2.39(0.29) & 5.54 & 2.05 & 298 & 23.71(0.23) & 17.40(0.28) & -6.32(0.23) \\
I18032-2032-HC5 & 271.56169 & -20.52703 & 2822 & 1866 & 67 & 24848(4713) & 0.49(0.04) & 7.70(0.56) & 346(289) & 2.11(0.22) & 5.32 & 0.75 & 231 & 23.52(0.17) & 16.89(0.23) & -6.64(0.15) \\
I18056-1952-HC1* & 272.15931 & -19.86399 & 9824 & 8830 & 177 & 31996(1821) & 1.22(0.06) & 7.70(0.03) & 2856(302) & 3.42(0.20) & 6.87 & 73.11 & 831 & 24.28(0.18) & 20.01(0.76) & -4.25(0.74) \\
I18089-1732-HC1 & 272.96441 & -17.52467 & 3231 & 1624 & 146 & 10325(456) & 0.87(0.03) & 7.52(0.10) & 698(189) & 1.94(0.15) & 5.26 & 2.02 & 461 & 23.95(0.15) & 17.89(0.50) & -6.09(0.56) \\
I18117-1753-HC1 & 273.66467 & -17.86671 & 3376 & 2011 & -175 & 23068(2631) & 0.49(0.03) & 7.73(0.11) & 406(97) & 2.15(0.14) & 5.08 & 1.10 & 237 & 23.53(0.20) & 16.95(0.22) & -6.62(0.17) \\
I18134-1942-HC1 & 274.09226 & -19.69092 & 2255 & 790 & -143 & 3302(79) & 0.42(0.01) & 8.41(0.06) & 192(28) & 2.75(0.18) & 3.14 & 0.25 & 141 & 23.61(0.24) & 15.95(1.11) & -7.95(0.31) \\
I18159-1648-HC1 & 274.72781 & -16.79733 & 3450 & 2096 & 81 & 10333(1541) & 0.66(0.05) & 8.10(0.11) & 557(200) & 2.97(0.60) & 4.80 & 2.31 & 304 & 23.78(0.25) & 16.93(0.82) & -6.97(0.85) \\
I18159-1648-HC3 & 274.72634 & -16.79716 & 1831 & 1085 & -176 & 8905(762) & 0.58(0.03) & 8.14(0.31) & 371(220) & 2.69(0.52) & 4.45 & 0.78 & 245 & 23.97(0.16) & 16.87(0.24) & -7.08(0.15) \\
I18182-1433-HC2 & 275.28800 & -14.53015 & 1702 & 1240 & -156 & 21251(5726) & 0.40(0.04) & 7.49(0.10) & 969(402) & 3.21(0.95) & 4.57 & 0.58 & 207 & 23.82(0.09) & 16.69(0.19) & -7.15(0.20) \\
I18236-1205-HC1 & 276.60746 & -12.06474 & 2090 & 1633 & 150 & 5207(1554) & 0.63(0.12) & 7.87(0.02) & 667(51) & 3.56(0.20) & 4.05 & 0.87 & 178 & 23.75(0.18) & 16.15(0.48) & -7.67(0.39) \\
I18264-1152-HC1 & 277.30980 & -11.83957 & 4953 & 3024 & -144 & 6414(608) & 0.67(0.05) & 7.84(0.30) & 232(153) & 1.53(0.12) & 4.46 & 2.79 & 207 & 23.69(0.17) & 16.28(0.99) & -7.48(1.05) \\
I18290-0924\_2-HC1 & 277.93384 & -9.37007 & 1684 & 1495 & 118 & 7209(607) & 0.83(0.05) & 7.41(0.04) & 1588(418) & 5.51(1.61) & 4.98 & 0.67 & 307 & 23.80(0.12) & 17.30(0.37) & -6.51(0.46) \\
I18316-0602-HC1 & 278.58717 & -5.99506 & 4656 & 3813 & 151 & 30318(11915) & 0.41(0.06) & 7.53(0.06) & 967(177) & 2.41(0.19) & 4.80 & 4.18 & 202 & 23.71(0.22) & 16.33(0.48) & -7.37(0.38) \\
I18411-0338-HC1 & 280.94263 & -3.59165 & 2975 & 2308 & 98 & 34133(6512) & 0.50(0.04) & 7.88(0.17) & 977(298) & 3.52(0.45) & 5.37 & 3.14 & 305 & 24.03(0.13) & 17.30(0.33) & -6.79(0.32) \\
I18434-0242-HC1 & 281.51574 & -2.65620 & 1599 & 1444 & 148 & 38158(12335) & 0.53(0.07) & 7.78(0.28) & 678(326) & 2.25(0.22) & 5.37 & 1.22 & 356 & 24.12(0.08) & 17.90(0.17) & -6.19(0.23) \\
I18469-0132-HC1* & 282.38772 & -1.48430 & 3544 & 2393 & 163 & 71620(13037) & 0.35(0.02) & 8.16(0.38) & 441(199) & 2.47(0.12) & 5.20 & 3.09 & 249 & 23.91(0.20) & 16.75(0.32) & -7.21(0.25) \\
I18479-0005-HC1 & 282.62805 & -0.03316 & 6725 & 4961 & 96 & 102640(52244) & 0.44(0.08) & 7.61(0.03) & 3391(356) & 3.21(0.21) & 5.66 & 41.87 & 262 & 24.48(0.10) & 16.88(0.17) & -7.61(0.20) \\
I18507+0110-HC1* & 283.32742 & 1.24947 & 3577 & 2886 & 120 & 44263(3156) & 0.55(0.02) & 7.42(0.08) & 4167(1767) & 5.26(2.51) & 5.67 & 4.86 & 406 & 24.10(0.35) & 17.97(0.68) & -6.24(0.86) \\
I18507+0110-HC2 & 283.32787 & 1.24929 & 1738 & 1481 & 51 & 50124(7044) & 0.40(0.02) & 7.92(0.04) & 1082(207) & 2.49(0.29) & 4.68 & 3.18 & 292 & 24.47(0.09) & 17.59(0.24) & -6.86(0.27) \\
I18507+0110-HC3 & 283.32796 & 1.24886 & 1563 & 1220 & 109 & 18775(9258) & 0.46(0.10) & 8.05(0.04) & 987(202) & 3.54(0.63) & 4.51 & 1.89 & 295 & 24.38(0.12) & 17.06(0.35) & -7.37(0.43) \\
I18507+0110-HC4 & 283.32781 & 1.24858 & 1748 & 1455 & 100 & 38078(22215) & 0.40(0.07) & 7.90(0.07) & 419(79) & 1.75(0.07) & 4.69 & 1.64 & 251 & 24.16(0.10) & 16.99(0.31) & -7.19(0.33) \\
I18507+0121-HC1 & 283.32504 & 1.42372 & 2261 & 1933 & 132 & 11130(1464) & 0.74(0.07) & 8.31(0.23) & 242(88) & 1.76(0.05) & 4.98 & 2.40 & 364 & 24.07(0.16) & 17.83(0.48) & -6.31(0.53) \\
I18517+0437-HC1 & 283.55933 & 4.69461 & 4488 & 1935 & -163 & 10366(1458) & 0.47(0.03) & 8.33(0.62) & 192(183) & 2.01(0.20) & 4.23 & 1.60 & 164 & 23.52(0.24) & 15.86(0.53) & -7.62(0.39) \\
I19095+0930-HC1* & 287.97500 & 9.59729 & 12427 & 9064 & 64 & 526045(414984) & 0.27(0.05) & 8.70(2.87) & 315(965) & 2.28(0.13) & 5.83 & 20.14 & 217 & 23.48(0.56) & 16.37(0.27) & -7.12(0.48) \\
\enddata
\tablenotetext{a}{An asterisk (*) denotes hot cores with detected H30$\alpha$ emission.}
\tablenotetext{b}{$\langle \rangle_{M}$ denote values mass-averaged over the hot core region in the corresponding map.}
\tablenotetext{c}{This represents the geometric center of the elliptical structure identified by \astrodendro{}, rather than the continuum peak position.}
\end{deluxetable*}
\end{longrotatetable}

\bibliography{main}{}
\bibliographystyle{aasjournalv7}

\end{document}